%% file: ms.tex
\documentclass[11pt,preprint]{emulateapj}
\usepackage{epsfig,natbib}
\usepackage[usenames]{color}
\def\redtext#1{#1}
\def\bluetext#1{#1}

\slugcomment{Submitted to the Astrophysical Journal Supplements on 10 June 2011}
\shortauthors{Eracleous et al.}
\shorttitle{Search for Binary Supermassive Black Holes}

\def\aj{\rm{AJ}}                   
\def\apj{\rm {ApJ}}                
\def\apjl{\rm{ApJ}}                
\def\apjlett{\rm{ApJ (Letters)}}   
\def\apjs{\rm{ApJS}}               
\def\aap{\rm{A\&A}}                
\def\aaps{\rm{A\&AS}}              
\def\mnras{\rm{MNRAS}}             
\def\nat{\rm{Nature}}              
\def\pasj{\rm{PASJ}}    	   
\def\na{\rm{New~Astr.}}            
\def\sovast {\rm{Sov.~Astr.}}      
\def\prd{\rm{Phys.\ Rev.\ D}}      
\def\prl{\rm{Phys.\ Rev.\ Lett.}}  
\def\pr{\rm{Phys.\ Rev.}}          

\def\cd{c$\!\!\!\hskip 0.75pt$\raise 0.2pt \hbox{\symbol{24}}}

\def\Msol{\ifmmode{\rm M}_{\mathord\odot}\else M$_{\mathord\odot}$\fi}
\def\Mbh{\ifmmode{M_{\rm bh}}\else{$M_{\rm bh}$}\fi}

\def\ls{\lower 2pt \hbox{$\;\scriptscriptstyle \buildrel<\over\sim\;$}} 
\def\gs{\lower 2pt \hbox{$\;\scriptscriptstyle \buildrel>\over\sim\;$}}

\def\kms{\ifmmode{\;{\rm km~s^{-1}}}\else{~km~s$^{-1}$}\fi}
\def\ergs{\ifmmode{\;{\rm erg~s^{-1}}}\else{~erg~s$^{-1}$}\fi}
\def\m#1{\ifmmode{^{-#1}}\else{$^{-#1}$}\fi}

\newcounter{species}
\def\ion#1#2{\setcounter{species}{#2}#1$\;${\sc\roman{species}}\relax}
\def\lion#1#2#3{\setcounter{species}{#2}#1$\;${\sc\roman{species}$\;\lambda${#3}}\relax}

\def\fion#1#2{[{\setcounter{species}{#2}#1$\;${\sc\roman{species}}\relax}]}
\def\flion#1#2#3{[{\setcounter{species}{#2}#1$\;${\sc\roman{species}}]$\;\lambda${#3}}\relax}
\def\fllion#1#2#3{[{\setcounter{species}{#2}#1$\;${\sc\roman{species}}]$\;\lambda\lambda${#3}}\relax}

\begin{document}

\title{A Large Systematic Search for Recoiling and Close Supermassive
  Binary Black Holes}

\author{Michael Eracleous\altaffilmark{1,2}, Todd
  A. Boroson\altaffilmark{3}, Jules P. Halpern\altaffilmark{4}, \&
  Jia Liu\altaffilmark{4}}

\altaffiltext{1}{Department of Astronomy and Astrophysics and Center
  for Gravitational Wave Physics, The Pennsylvania State University,
  525 Davey Lab, University Park, PA 16803, U.S.A.}

\altaffiltext{2}{Visiting Astronomer, Kitt Peak National Observatory,
  National Optical Astronomy Observatory, which is operated by the
  Association of Universities for Research in Astronomy (AURA) under
  cooperative agreement with the National Science Foundation.}

\altaffiltext{3}{National Optical Astronomy Observatory, Tucson, AZ
  85719, U.S.A.}

\altaffiltext{4}{Columbia Astrophysics Laboratory, Columbia
  University, 550 West 120th Street, New York, NY 10027-6601, U.S.A.}


\begin{abstract}
We have carried out a systematic search for {\it close} supermassive
black hole binaries (with sub-parsec separations) among $z\ls 0.7$
quasars observed spectroscopically in the Sloan Digital Sky
Survey. Such binaries are predicted by models of supermassive black
hole and host galaxy co-evolution, therefore their census and
population properties constitute an important test of these models.
\bluetext{Our working hypothesis is that one of the two black holes
  accretes at a much higher rate than the other and carries with it
  the only broad-emission line region of the system, making the system
  analogous to a single-lined spectroscopic binary star.} Accordingly,
we used an automatic technique based on spectroscopic principal
component analysis to search for broad H$\beta$ emission lines that
are displaced from the rest-frame of the quasar by $|\Delta v| \gs
1,000\;{\rm km\; s^{-1}}$ (corresponding to binary periods and
separations of $P\sim {\rm few} \times 100\;$yr and $a\sim {\rm few}
\times 0.1\;$pc, respectively, for masses $\sim 10^8\;{\rm
  M}_\odot$). This method can also yield candidates for rapidly
recoiling black holes since their spectroscopic signature is
similar. Our search yielded 88 candidates, \redtext{several} of which
were previously identified and discussed in the literature. The widths
of the broad H$\beta$ lines are typical among quasars but the shifts
are extreme. We found a correlation between the peak offset and
skewness of the broad H$\beta$ profiles (there is an extended wing on
the opposite side of the profile from the shifted peak), which
suggests that the profiles we have selected \bluetext{share a common
  physical explanation}. The general properties of the narrow emission
lines are typical of quasars.  We carried out followup spectroscopic
observations of 68 objects to search for changes in the peak
velocities of the H$\beta$ lines (the time interval \redtext{in the
  observer's frame} between the original and new observations is
1--10~yr and 5.7--10.0~yr in 2/3 of the cases). We measured
statistically significant changes (at 99\% confidence) in 14 objects,
with resulting accelerations between \redtext{$-120$ and
  +120~km~s$^{-1}$~yr$^{-1}$}.  The above results, taken at face
value, are broadly consistent with predictions for the number of close
supermassive binaries in the Sloan Digital Sky Survey quasar
sample. \bluetext{However, such a comparison is complicated by several
  theoretical and observational uncertainties, such as the fact that
  the observable we employ to select objects depends on a combination
  of several degenerate intrinsic parameters of a binary. We emphasize
  that interpretation of the offset broad emission lines as signatures
  of supermassive binaries is subject to many significant
  caveats. Many more followup observations over a long temporal
  baseline are needed to characterize the variability pattern of the
  broad lines and test that this pattern is indeed consistent with
  orbital motion.  The possibility that some of the objects in this
  sample are rapidly recoiling black holes remains open as the
  available data do not provide strong constraints for this scenario.}
\\ \\ 
{\bf All tables and two large figures in this preprint have been
  abridged. A version with all the figures and a complete list of the
  sample objects can be found at:} 
\\ \\
\centerline{{\tt http://www2.astro.psu.edu/users/mce/preprints/SBHB.pdf} (5 MB)}
\\
\end{abstract}



\section{Introduction}\label{S:intro}

\subsection{Theoretical Background and Motivation}\label{S:intro-theory}

The realization that the masses of the nuclear supermassive black
holes (BHs) hosted by massive galaxies are related to the mass of the
stellar spheroid of the host \citep[e.g.,][]{ferrarese00, gebhardt00}
has led to the development of new numerical and semi-analytic galaxy
evolution models. Such models \citep[e.g.,][]{volonteri03, dimatteo05,
  hopkins06} attribute the observed correlation to co-evolution of the
black hole and the host via hierarchical galaxy mergers and accretion.
Close supermassive black hole binaries (SBHBs) appear to be an
inevitable stage in the evolution of the post-merger remnant. SBHBs
have also been invoked to explain a variety of observations. Examples
include stellar dynamical models that seek to explain the formation of
cores in elliptical galaxies following a merger and mass deficits
therein \citep[e.g.,][]{milosavljevic02, merritt06a}, and the
morphologies of some radio sources \citep*[such as precessing jets and
  X-shaped sources; see][]{roos93, romero00, merritt02,
  gopal-krishna03}.  SBHBs are also expected to be among the primary
sources of gravitational waves detectable by the Laser Interferometer
Space Antenna \citep[LISA; e.g.,][]{sesana05}.

The evolution of SBHBs following the merger of their parent galaxies
was first described by \citet*{begelman80}. The early evolution
(sinking of the BHs towards the center of the remnant and decay of the
orbit via dynamical friction) proceeds relatively quickly, on a time
scale comparable to the dynamical time of the parent galaxies ($\sim
10^8$ years). Once the binary ``hardens'' (i.e., the orbital speed
becomes comparable to the stellar velocity dispersion) the evolution
slows down as the orbital decay relies primarily on scattering of
stars. If the number of stars whose orbits allow close encounters with
the binary can be replenished efficiently, the process can shrink the
orbit to a separation where the emission of gravitational waves
becomes an effective orbital decay mechanism and the binary can
coalesce within a Hubble time. However, it is still unknown whether
this replenishment is possible via stellar dynamical processes
\citep*[see, for example, discussions in][]{milosavljevic03,
  berczik06, sesana07}. Therefore, the possibility that the orbital
evolution of the binary may stall at a separation $\ls 1\;$pc
remains open (this is known as ``the last parsec problem'').
Alternative mechanisms for the decay of the binary orbit have been
explored in an effort to solve this problem, most notably drag by a
large gaseous reservoir (e.g. a disk) or torques on a binary at short
separations by a circumbinary disk\citep[e.g.][]{escala04, dotti07,
  dotti09b, cuadra09, lodato09}.

Despite their importance and the theoretical attention they have
received, robust, {\it dynamical} evidence for the existence of large
numbers of SBHBs at separations $ <1\,$pc \cite[the slowest phase in
  their evolution according to][]{begelman80} remains
elusive. Discovering such systems and obtaining a census of their
population properties would serve as a test of galaxy evolution models
and would provide valuable constraints for stellar and gas dynamical
models for the decay of the binary orbit. It would also validate
explanations of observed phenomena that appeal to such SBHBs. 

The above evolutionary scenarios along with the new observational work
on this subject (described below) have prompted the development of
models for the observational signature of such binaries, such as the
behavior of the emission lines from gas accreting onto the two BHs.
The observational signature that is thought to provide the most
direct, dynamical evidence for SBHBs is based on an analogy with
spectroscopic binary stars. In this picture the emission lines from
gas bound to one or both of the BHs (the broad-line region or BLR) are
Doppler-shifted according to the line-of-sight velocity of the source.
\citet{bogdanovic08} have studied the interaction of the less massive
(hereafter secondary) BH with the accretion disk around the more
massive (primary) one in a close, eccentric binary and have
highlighted the potentially complex variability of the emission line
profiles. \citet*{hayasaki07} and \citet{cuadra09} developed
smoothed-particle hydrodynamic (SPH) models for the gas flow from a
circumbinary disk onto disks around the individual BHs, based on which
they make predictions about the relative accretion rates.  In the
context of such a picture of a binary within a circumbinary disk,
\citet{montuori10} have used photoionization models to compute the
relative strengths of the broad emission lines from these systems in
an effort to devise spectroscopic diagnostics for
SBHBs. \citet{shen10a} present a toy model of the BLRs in SBHB in
which they employ test particles to represent clouds of gas in
gravitationally bound orbits in the system and discuss its
consequences for observational searches for SBHBs.  Although this
model is unrealistic\footnote{In addition to the shortcomings
  mentioned by the authors themselves, we also note that the basic
  premise of a BLR consisting of discrete clouds is not supported by
  observations \citep[see, for example,][and references
    therein]{laor06}. Even if a spherical system with crossing orbits
  like the one assumed could be set up, it would be destroyed by
  collisions on a very short time scale \citep[see][]{mathews85}.}
and the predicted line profiles cannot be used as a basis for
quantitative tests, the authors make the very important point that, at
short separations the BLR would envelope the entire binary. This point
was also made by M. Penston in 1988, as noted in footnote 3 of
\citet*{chen89}.

A related development in numerical relativity is the calculation of
the recoil speed (or ``kick'') of a BH resulting from the inspiral and
merger of two BHs. A review of the recent literature on this subject
can be found in \citet{centrella10a, centrella10b}. The most extreme
recoil speed expected from this effect is $\sim4000\kms$ for a special
orientation of the spins of the two BHs relative to their orbital
plane. However, under most other configurations, the recoil speed is
considerably lower than this value \citep[e.g.,][]{ campanelli07a,
  campanelli07b, baker08}. As noted by \citet*{bogdanovic07}, in a
realistic astrophysical case the spins of the two BHs may be aligned
with the orbital axis as a result of the previous evolution of the
binary in a gaseous environment. In such a case, the recoil speeds
would be rather low, namely of order a few hundred\kms. Indeed,
\citet{dotti10} predict, based on detailed and extensive simulations,
that the recoil speeds are likely to be distributed in the range
10--100\kms.

As in the SBHB case, finding recoiling BHs and studying their
population properties would be quite valuable. It would provide a test
of calculations for the recoil speed but also much-needed constraints
on models of the astrophysical implications for such kicks. A
particularly interesting consequence of kicks is that they can propel
the BH out of the host galaxy if they are strong enough (i.e., $\gs
1,000$\kms), which could influence the cosmological evolution of BHs
\citep[e.g.,][]{volonteri10}. Otherwise, kicks can disturb the
morphology of the host galaxy or offset an active nucleus from the
center of its host \cite[e.g.,][]{gualandris08, komossa08b, blecha10}.
A potential observational signature of recoiling BHs is a Doppler
shift of the broad emission lines originating in a BLR that is bound
to the recoiling BH \citep[see][]{loeb07}. Another possibility is that
the recoiling BH may accrete gas that it encounters along its
trajectory, which could lead to a prolonged AGN phase lasting up to
0.1--1~Gyr \citet{blecha10}.  For large kicks, the spectroscopic
signature of such a system is similar to the signature one might
expect from some sub-parsec SBHBs, implying that a systematic search
for SBHBs based on this idea may also yield useful constraints on the
frequency of large kicks.

\subsection{Observational Searches for Close Binary and Recoiling 
Holes in Historical Perspective}\label{S:intro-obs}

Until recently, direct imaging had resolved very few (presumably
unbound) wide BH pairs, such as NGC\,6240 \citep{komossa03}, and a
close pair, CSO\,0402+379 \citep{maness04, rodriguez06,
  rodriguez09}. Recent spectroscopic surveys, however, have uncovered
a substantial population of moderate-redshift AGNs with offset or
double-peaked, narrow \fion{O}{3} emission lines. Although some of the
offset or double-peaked \fion{O}{3} lines could arise in biconical
outflows from single active nuclei \citep[see][]{crenshaw10},
some of them turn out to trace dual active nuclei in merging galaxies,
as shown by followup studies \cite[e.g.,][]{comerford09a,
  comerford09b, liu10a, liu10b, shen10b, smith10b, fu10}. These dual
active nuclei have separations of order a few kiloparsec \citep[a
  representative example is COSMOS\,J100043.15+020637.2, reported
  by][]{comerford09b}.

In contrast, close, bound binaries, at sub-parsec separations, have
been quite elusive as they cannot be resolved spatially. Nevertheless,
such close binaries are of particular interest because they represent
the slowest phase in the evolution of such a system, according to
\citet{begelman80}. In a well-known case, the $\sim 12$-year periodic
modulation in the long-term light curve of OJ~287 has been explained
in terms of a close SBHB \citep[e.g.,][and references
  therein]{valtonen08}. Other cases rely on a different type of
indirect evidence, the detection of broad emission-lines with
displaced peaks in quasar spectra. The underlying hypothesis here is
that one or both of the two BHs in a SBHB has a gaseous BLR associated
with it that emits the broad, permitted lines that are characteristic
of quasar spectra. Thus, the orbital motion of the binary causes the
lines to shift periodically making the spectrum analogous to that of a
single- or double-lined spectroscopic binary star
\citep[see][]{komberg68, begelman80}.  In the context of this
hypothesis, \citet{gaskell83, gaskell84} identified two quasars,
3C~227 and Mrk~668, whose broad H$\beta$ emission lines were shifted
from their nominal positions by 2,000--3,000\kms\ and proposed that
these are examples of close SBHBs in which only one of the two BLRs is
visible. He also suggested that quasars with double-peaked emission
lines, such as 3C~390.3, could represent SBHB in which both BLRs are
visible. Indeed, an observed systematic drift of the blue-shifted,
broad H$\beta$ peak in the spectrum of 3C~390.3 between 1968 and 1988
was consistent with this hypothesis \citep{gaskell96}. However, this
behavior is a rather uncommon feature of double-peaked emission lines
\citep[e.g.,][]{gezari07,lewis10} and, in fact, the trend observed in
3C~390.3 did not continue past 1990 \citep{eracleous97,
  shapovalova01}. Moreover, in a detailed study of the long-term
variability of double-peaked emission lines in three quasars,
\citet{eracleous97} did not detect the behavior expected from a binary
BLR associated with an SBHB and also noted a number of contradictions
between this interpretation and the observed properties of these
objects. Finally, long-term variability studies of about two dozen
double-peaked emitters \citep{gezari07,lewis10} do not show the
signature of orbital motion and a number of arguments suggest that
broad double-peaked lines are much more likely to originate in an
accretion disk \citep[see discussion in][]{eracleous03}. Nevertheless,
the binary black hole hypothesis remains a reasonable explanation for
broad, {\it single-peaked} Balmer lines that are shifted from their
nominal wavelengths. It is interesting that a number of such cases
have been known for quite some time among nearby Seyfert galaxies. An
inspection of the spectroscopic atlas of \citet{stirpe90} reveals
several examples, of which Mrk~876 and NGC~5548\,\footnote{The
  variations of the H$\beta$ profile of NGC~5548 were discussed by
  \citet*{peterson87} in the context of a SBHB model.  However, the
  wealth of data available today indicate that the variability
  properties of this object do not support such an interpretation
  \citep[e.g.,][]{wanders96, peterson02, sergeev07}.} are particularly
striking.

The recent theoretical work on the evolution of black holes in merging
galaxies has been accompanied by new observational work directed at
identifying close (sub-parsec) SBHBs. A handful of quasars from the
Sloan Digital Sky Survey (SDSS), whose broad H$\beta$ lines have peaks
that are displaced from their nominal wavelengths by a few
thousand\kms, have been noted and discussed. The quasar
SDSS~J092712.65+294344.0 was interpreted as a SBHB by
\citet*{bogdanovic09} and \citet{dotti09a}, although it was proposed
by \citet*{komossa08a} as a candidate recoiling BH (see also our
discussion below). The quasar SDSS~J153636.22+044127.0 was identified
by \citet{boroson09} who attributed its strong, broad, blueshifted
H$\beta$ peak to a BLR around one of the members of a SBHB. However,
\citet{chornock10} and \citet{gaskell10} called this interpretation
into question based on the presence of a prominent red shoulder in the
H$\alpha$ profile, which is also discernible in the H$\beta$ profile,
and suggested that the Balmer lines originate in a perturbed disk
around a single black hole. Yet another interpretation for this object
was offered by \cite{tang09}, who proposed that this quasar does
harbor a SBHB and the profiles of the Balmer lines comprise
contributions from an accretion disk around the primary BH, a BLR
associated with the secondary BH, and a more extended gaseous region
surrounding the binary. An interpretation similar to that of
\citet{tang09} was also discussed by \citet{barrows11} for the
H$\beta$ profile of SDSS~J093201.60+031858.7. A fourth case, 4C+22.25
(a.k.a. SDSS~J100021.80+223318.7) was recently reported and discussed
by \citet{decarli10}. The H$\beta$ profile of this object is
reminiscent of SDSS~J092712.65+294344.0 (a broad peak blueshifted by
8,700\kms), but the large blueshift of the peak makes a
rapidly-recoiling BH interpretation untenable.  \redtext{In a very
  recent study, \citet{tsalamantza11} have used an automatic method to
  search systematically through the SDSS DR7 spectroscopic database
  (including 54,586 objects classified as quasars and 3,929 objects
  classified as galaxies at $0.1<z<1.5$) for objects with displaced
  emission line peaks. They have uncovered five new SBHB candidates,
  in addition to the ones noted above.}  All of the above objects
should be regarded as preliminary candidates for SBHBs in the absence
of additional evidence. This is because our lack of understanding of
the structure and properties of the BLR does not preclude that such
unusual line profiles can originate from the BLR around a single BH.
The case would be strengthened if the displaced peaks are observed to
drift in time, as one would expect for the lines of a spectroscopic
binary star.

A systematic search for recoiling BHs was undertaken only recently,
motivated by the new developments in numerical
relativity. \citet{bonning07} studied the distribution of shifts of
the broad H$\beta$ lines in the SDSS DR3 quasar sample and found that
shifts greater than 800\kms\ occur in only 0.2\% of cases, with
progressively more stringent limits for larger shifts. The latest
studies for the distribution of kicks \citep[e.g.,][]{dotti10}
suggest that the vast majority of recoil speeds should be $\ls
100\kms$, which means that recoiling BHs would not be easily
identifiable in the sample of \citet{bonning07}.

Three candidate {\it rapidly} recoiling black holes have been noted
and discussed in the recent literature, but their cases are ambiguous
or controversial.  As we mentioned above, SDSS~J092712.65+294344.0 was
identified by \citet{komossa08a} via its shifted H$\beta$ line, who
suggested that this is a rapidly-recoiling black hole. However, a
number of different interpretations were offered by followup papers,
including a binary black hole \citep{bogdanovic09, dotti09a}, a chance
superposition of objects at different redshifts on the sky
\citep{shields09b}, and a superposition of objects in the same,
massive galaxy cluster \citep{heckman09}.  Similarly
SDSS~J105041.35+345631.3 was discovered through a visual inspection of
a large number of SDSS quasars and discussed by \citet{shields09a} and
\citet{smith10a}. These authors noted that the relatively symmetric
displaced peaks of the Balmer lines suggest a rapidly recoiling BH but
differences between the profiles of broad lines from different ions
(\ion{H}{1} and \ion{Mg}{2}) suggests that the lines are actually
double-peaked but extremely asymmetric, which would point to an origin
in a perturbed accretion disk. Finally, \citet{civano10} suggested
that one of the two nuclei CID-42 (a.k.a. COSMOS\,J100043.15+020637.2;
see the first paragraph of this section) could also be a rapidly
recoiling BH.

The ambiguity surrounding the objects mentioned in the previous
paragraph stems from the fact that the available data do not allow us
to definitively distinguish between SBHBs and rapidly recoiling BHs.
The detection of a change in the velocity offset of the broad lines
would favor the SBHB over the recoiling BH interpretation, although it
would not provide iron-clad evidence, as we discuss in more detail in
later sections of this paper. One can argue against a recoiling BH
interpretation in some particular cases on theoretical grounds, since
recoil speeds far in excess of 100\kms\ are extremely unlikely
\citep[e.g.,][]{dotti10} and speeds above 4,000\kms\ are unattainable
\citep[e.g.,][]{campanelli07a, campanelli07b}. An indirect test has
been described by \citet{bonning07} and re-iterated by
\citet{shields09a} in which the displaced broad lines have fairly
symmetric profiles and the prominent narrow lines have similar widths,
regardless of ionization state.  The underlying hypothesis here is that
the broad lines are emitted from an otherwise ordinary BLR that is
attached to the recoiling BH \citep[see][]{loeb07} and that the narrow
lines are emitted from the galaxy where the BH originated, which is
now illuminated from the outside.

\section{Scope and Goals of Our Program and Considerations Underlying 
the Selection of Candidates and Search Strategy}\label{S:design}

Motivated by the above, we have embarked on a systematic search for
quasars whose broad (single-peaked) H$\beta$ lines are offset from
their nominal wavelengths by $|\Delta v| \gs 1,000\kms$. We regard
such objects as candidates for close (sub-parsec) SBHBs or rapidly
recoiling BHs and have begun followup observations to search for
changes in the shifts of their broad lines. Even though we adopt this
as our working hypothesis, we emphasize that such line profiles need
not be signposts of SBHBs. One can envision scenarios in which such
line profiles can originate from the BLR associated with a single BH
that does not have a companion.  Thus, the mere detection of an offset
in the broad Balmer lines yields only weak candidates for SBHBs or
rapidly recoiling BHs. The detection of gradual drifts in the broad
Balmer lines via followup observations strengthens the case for a
SBHB, but a convincing demonstration that a specific object harbors a
SBHB requires us to observe (ideally) a few orbital cycles. We return
to these issues in \S\ref{S:discussion}, where we offer a more
extensive discussion.

Based on the models of \citet{volonteri09} we would optimistically
expect up to $\sim 160$ sub-parsec SBHBs in the sample that we have
searched. However, there are significant theoretical uncertainties and
models by these authors based on different assumptions yield a number
that is lower by an order of magnitude.  Since the SBHB phase is
thought to be considerably longer than the accretion phase of a
recoiling BH and since the expected distribution of recoil speeds is
heavily weighted towards low speeds ($|v| \ll 1,000\kms$) we expect
that only a very small fraction of the the quasars we select will be
candidate recoiling BHs \citep[see][]{bogdanovic07, dotti09a,
  dotti10}.

If the binaries evolve within a large-scale gaseous disk, as suggested
by recent theoretical studies \citep[e.g.,][]{dotti07, dotti09b}, we
would expect that their orbits will have circularized by the time they
reach sub-parsec orbital separations.  At this evolutionary stage,
binaries open a hole at the center of the circumbinary disk and the
secondary BH orbits closer to the gas reservoir and has easier access
to it. Therefore, the accretion rate onto the secondary is
considerably higher than that onto the primary (see the recent work by
Hayasaki et al. 2007 and Cuadra et al. 2009, building on previous
calculations by Artymowicz \& Lubow 1996 and Gould \& Rix 2000), which
leads us to attribute any shifted emission lines we detect to a BLR
associated with the secondary. The resonant interaction of the binary
with the circumbinary disk could lead to an increase in the
eccentricity. This is a somewhat uncertain process, which depends on
the thermodynamic properties of the circumbinary disk and the amount
of gas accreted by the BHs \citep[see][]{artymowicz91, armitage05,
  cuadra09, rodig11}.  \bluetext{We consider the accreting secondary
  scenario to be more likely on theoretical grounds, but our technique
  does not preclude us from finding systems in which the BLR is
  associated with the primary \citep[e.g.][]{chang10}, or systems with
  a BLR that envelops the entire binary.}

Since the main observable we employ is the velocity offset of the
broad H$\beta$ line, we consider here how the properties of detectable
binaries might depend on this quantity. For the sake of simplicity we
assume that the orbit is circular. Following \citet{bogdanovic09}, we
write the magnitude of the observed velocity shift (i.e., projected
along the line of sight) of the broad lines from the BLR of the
secondary, $u_2$, in terms of the true orbital speed, $V_2$, as $u_2=
V_2\,\sin i\,|\sin\phi|$, where $i$ is the inclination of the orbit
($i=0^\circ$ is face on), $\phi=2\pi(t-t_0)/P$ is the orbital phase,
and $P$ is the period.  We express the period and orbital separation
of the binary in terms of the total mass, $M_8=(M_1+M_2)/10^8\;{\rm
  M_\odot}$, mass ratio, $q=M_2/M_1 < 1$, and the projected velocity
of the secondary, $u_{2,3}=u_2/10^3\kms$ as:
\begin{equation}
P=2652\, M_8\, \left[\sin i\,|\sin\phi|\over(1+q)\,u_{2,3}\right]^3~{\rm yr}
\label{Q:period}
\end{equation}
and
\begin{equation}
a=0.432\,M_8\, \left[\sin i\,|\sin\phi|\over(1+q)\,u_{2,3}\right]^2~{\rm pc.}
\label{Q:separation}
\end{equation}
If the emission lines trace the motion of the the primary instead, the
above equations can be adapted by replacing $u_{2,3}$ by $u_{1,3}/q$,
where $u_{1,3}=u_1/10^3\kms$ ($u_1$ is the projected speed of the
primary). For the sake of estimating typical binary periods and
separations, we set $i=\phi=45^\circ$, which yields 
\begin{equation}
P={332\; M_8 \over (1+q)^3\, u_{2,3}^3}
\left({\sin i \over \sin 45^\circ}\,{|\sin\phi|\over\sin 45^\circ}\right)^3
~{\rm yr} 
\label{Q:periodnum}
\end{equation}
and
\begin{equation}
a={0.11\,M_8\over(1+q)^2\, u^2_{2,3}}
\left({\sin i \over \sin 45^\circ}\,{|\sin\phi|\over\sin 45^\circ}\right)^2
~{\rm pc}.
\label{Q:separationnum}
\end{equation}
In other words, if we measure the projected speed of the secondary BH
to be of order $10^3\kms$ in a binary with $q\ll 1$ and $M_8\sim 1$,
then it is reasonable to expect the binary to have a period of order a
few centuries and a sub-parsec separation. We note here that the mass
of the secondary cannot be arbitrarily small since it is constrained
by the requirement that the accretion luminosity be high enough for
this object to appear as a quasar.  If the emission lines trace the
motion of the primary instead, then the period and separation would be
smaller by factors of $q^3$ and $q^2$, respectively. These long
orbital periods highlight a significant difficulty in securely
identifying SBHBs: orbital periods are very long compared to a human
lifetime, hence the observation of even a single orbital cycle
requires considerable patience and persistence.  In an optimistic
scenario, where the binary has a mass and a projected velocity offset
such that $M_8\approx 0.1$ and $u_{2,3}\approx 2$, the orbital period
could be as short as a decade, however. We will return to this issue
in \S\ref{S:discussion} where we will consider the likelihood of
observing a system with a specific set of properties.

As we noted above, the case for a displaced broad line signaling the
presence of a binary becomes stronger if the shift of the line is
observed to change with time. In the context of the binary picture
outlined in the previous paragraph, the projected acceleration is
given by
\begin{eqnarray}
\left|du_2\over dt\right|
& = & 2.4\;{u^4_{2,3}\,(1+q)^3\over M_8\,\sin^3i}\, 
\left|\cos\phi\over\sin^4\phi\right| \kms\;{\rm yr}^{-1} 
\nonumber \\
\label{Q:accelerationnum} \\
& = & 19\;{u^4_{2,3}\,(1+q)^3\over M_8}\; 
      \left(\sin 45^\circ\over \sin i\right)^3  \nonumber \\
&   & \times \, 
{|\cos\phi|\over\cos 45^\circ}
\left(\sin 45^\circ \over \sin\phi\right)^4 \kms\;{\rm yr}^{-1}
\nonumber \\
\nonumber 
\end{eqnarray}
The time intervals \redtext{(in the observer's frame)} between the
original SDSS observations and the new observations carried out in our
program are in the range of 5--10 years. Thus, for sub-parsec binaries
with orbital periods of order centuries, it is reasonable to expect
the line shifts to have changed by 100--200\kms. Our survey was,
therefore, designed to detect changes of this order. If, however, the
broad line shifts trace the motion of the primary, then,
\bluetext{{\it for a given observed velocity} the accelerations are
  larger by a factor of $q^{-3}$ compared to the expression in
  equation~(\ref{Q:accelerationnum}) and are easier to detect.}

It is noteworthy that $|du_j/dt|\propto V_j\,|\cos\phi|$, while
$u_j\propto V_j\,|\sin\phi|$ (where $j=1,2$ labels the BH), i.e., the
speed and acceleration are $1\over 4$ cycle out of phase during their
oscillation. This means that binaries near conjunction should display
small velocity offsets but large accelerations, while binaries near
\redtext{quadrature} should display large velocity offsets and small
accelerations. As noted by \citet{bogdanovic09}, this represents an
inherent difficulty in making a case for a SBHB because large
accelerations should be accompanied by small line shifts and vice
versa.

With these considerations in mind, in \S\ref{S:selection} we describe
in detail how we selected our sample out of the SDSS spectroscopic
database and we present the basic properties of the resulting objects,
while in \S\ref{S:sample-properties} we present a detailed study of
the spectroscopic properties of these objects. In \S\ref{S:newobs} we
present our followup (``2nd-epoch'') observations and compare the
followup spectra to the SDSS spectra.  The techniques we used to
search for velocity changes in the displaced broad peaks are described
in \S\ref{S:shifts} and in \S\ref{S:discussion} we summarize and
discuss our results.

\section{Target Selection}\label{S:selection}

The basis for our identification of candidates for the target list is
a principal component analysis (PCA) technique described in detail by
\citet{boroson10}.  We applied this technique to the SDSS spectra of
15,900 QSOs having redshift $z<0.7$, from the catalog of
\citet{schneider10}.  We note that this catalog includes only objects
more luminous than $M_i = -22$.  The input variables were the spectra
themselves, shifted to rest and trimmed to the wavelength range
4290--5400~\AA, centered roughly on the H$\beta$ region.  Eigenspectra
were generated from a subset of these (1,000 objects) with the highest
signal-to-noise.  Although the spectrum of almost every object could
be modeled to within its noise by using a large number of
eigenspectra, our goal was to identify the objects with the worst fits
as candidates for our sample.  Thus, we reconstructed each of the
15,900 spectra using only the five most significant eigenspectra.
This effectively limits the reconstruction such that it can only
reproduce the most common features and characteristics of the line
profiles.

We adopted a threshold of $\chi^2/\nu > 3$ (where $\nu $ is the number
of degrees of freedom, i.e., the number of spectral pixels) as a
starting point for identifying spectra that were fitted poorly and
were therefore candidate offset-peak objects.  This gave us 910
objects.  Note that this threshold value is arbitrary and is dependent
on the signal-to-noise ratio ($S/N$) of the spectrum.  A very
high-$S/N$ spectrum with a slightly unusual line profile will result
in a poor fit.  However, we thought that this approach would tend to
err on the side of including many objects that we would later reject,
and so it is inherently conservative.

We visually inspected the 910 spectra identified by the automatic
selection algorithm and applied the following additional criteria.  We
excluded most objects in which the peak of the broad H$\beta$ line was
shifted by less than 1,000\kms\ from the peak of the narrow H$\beta$
line (interpreted as the systemic redshift) since any offset
broad-line peak is blended with the narrow H$\beta$ line and difficult
to discern. Identifying the peak of the broad line was subjective in
some cases, particularly in noisy spectra.  We excluded H$\beta$
profiles with shelves and inflections, if they had a dominant broad
peak at $v\approx 0$ \citep[examples of quasars with such profiles
  that were excluded are SDSS J121716.08+080942.0,
  SDSS~J143452.45+483942.8, SDSS J154348.62+401324.9, and
  SDSS~J154929.43+023701.1; see Figure 5 of][]{zamfir10}. We also
excluded clear double-peaked emitters, i.e., objects with two clearly
separated broad peaks on either side of the narrow H$\beta$ line. This
is because variability studies have tested and rejected the hypothesis
that such profiles are indicative of SBHBs
\citep{eracleous97,gezari07,lewis10}. However, we retained objects
whose broad H$\beta$ profiles have a strong displaced peak plus an
extended wing or shoulder on the opposite side of the narrow H$\beta$
line \citep[a prime example of such an object is
  SDSS~J153636.22+044127.0, presented by][while additional examples
  from our collection are identified in
  \S\ref{S:demographics}]{boroson09}\footnote{Objects with such
  profiles have been termed ``extreme double-peaked emitters'' and
  have been discussed as alternative interpretations for broad
  displaced peaks to SBHBs \citep[e.g.][]{shields09b, lauer09,
    decarli10}. These two explanations are not mutually exclusive, of
  course. The unusual profiles of extreme double-peaked emitters could
  be the result of perturbation of the line-emitting disk from a
  companion BH in a SBHB \citep[e.g.,][]{eracleous95, bogdanovic08,
    tang09}. However, similar perturbations can be caused by the self
  gravity of the disk.\label{N:extreme}}. We also retained flat-topped
profiles, if these were not centered at $v=0$ since they could be a
manifestation of a single broad-line region bound to a SBHB
\citep[see][]{shen10a}.  This subjective, visual classification
process resulted in a sample of 88 objects that met all our criteria.
To these we added two more objects listed by \citet{strateva03}, which
had not been selected by our $\chi^2/\nu > 3$ condition.  Again, we
stress that this is not a complete sample in any sense, though we
believe that it includes almost all of the objects that fit our
criteria.

According to \citet{volonteri09} there may be up to 150 detectable
SBHBs with $q>0.01$ in the sample of 15,900 quasars we have
searched. However, models based on different assumptions predict up to
an order of magnitude lower numbers, and the predictions are, in fact,
consistent with zero when theoretical uncertainties are considered.

\input  tab1.tex
\input  tab2.tex

The 88 objects making up the resulting sample, along with their basic
properties are listed in Table~\ref{T:obslog}. This sample is
considerably larger than the few objects with case studies published
so far. In columns 1--5 of this table we give the redshift, apparent V
magnitude, Galactic V-band extinction in the direction of the source,
and the absolute V magnitude for each quasar. These quantities were
obtained as described in \S\ref{S:basic-properties}. Hereafter, we
will adopt abbreviated designations for these objects using the first
six digits of the right ascension (i.e., down to integer seconds, e.g.,
Jhhmmss). With this convention we can refer to the objects in
Table~\ref{T:obslog} uniquely. In columns 6--10 of
Table~\ref{T:obslog} we list the particulars of the spectroscopic
observations of each target, starting with the original SDSS
observations (additional rows refer to followup observations described
in \S\ref{S:newobs}). 

Some of the objects in Table~\ref{T:obslog} were identified previously
and discussed by other authors, as follows.

\begin{itemize}

\item
The quasar J140700 is Mrk~668, one of the first two candidate SBHBs
proposed by \citet{gaskell83}.  It is noteworthy that the other
candidate SBHB of \citet{gaskell83}, 3C~227, was also observed during
the SDSS (designated SDSS~J094745.14+072520.5) but was not selected by
our algorithm. This is because the profile of its broad H$\beta$ line
at the time of the SDSS spectroscopic observation (UT 2003/03/26) was
relatively symmetric and centered at the nominal wavelength of the
line. We will return to these two objects in our discussion of
interesting cases in later sections of the paper.

\item
We have recovered three objects that have received attention in the
recent literature: J092712 and J105041 (see \S\ref{S:intro-obs}), as
well J153636 (originally discovered by this method). We have also
recovered an object noted by \citet{bonning07},
J091833\footnote{\citet{bonning07} drew attention to this object
  because it had the largest velocity offset in their sample}. In
\S\ref{S:sdss-lines} we compare in more detail the properties of the
broad H$\beta$ lines from our sample to the findings of
\citet{bonning07} and to statistical studies by other authors.

\redtext{\item Our sample has 13 objects in common with the list of 32
  objects of \citet{tsalamantza11}, who also used an automatic
  selection algorithm to search through the SDSS DR7 spectroscopic
  database. Of these 13 objects, four are the objects listed in the
  previous paragraph, two are newly selected SBHB candidates by both
  us and them (J115449 and J171448), while the remaining seven are
  objects that we regard as SBHB candidates while they have classified
  as having asymmetric line profiles or ``other.'' There are four SBHB
  candidates in the list of \citet{tsalamantza11} that are not in our
  sample: J0932+0318, J1000+2233, J1012+2613, and J1539+3333 (using
  their naming convention). The second and third of these four they
  also classified as double-peaked emitters (we agree with this
  classification) which is why they are not included in our sample.}

\item
Five of the objects in our sample, J001224, J015530, J094603, J114755,
and J120924 were noted by \cite{zamfir10} as unusual cases that
illustrate the wide diversity of Balmer line profiles in quasars.
Several other objects in the \cite{zamfir10} study are worthy of
mention, because they are not included in our sample.  Two objects,
J101912 and J162345, are below the luminosity threshold of the
\citet{schneider10} QSO list.  Three objects, J074948, J084203, and
J143511, have broad H$\beta$ lines with peaks that are less than
1,000\kms\ from the narrow H$\beta$ line.  One of these appeared in
our candidate list with a $\chi^2/\nu$ value of 3.92 and was rejected
upon visual inspection.  The other two had $\chi^2/\nu$ values just
below our cutoff.

\item
Nine of the objects in our sample are also in the sample of $z< 0.33$
quasars with double-peaked H$\alpha$ lines by \cite{strateva03},
selected from the SDSS DR3. These are J001224, J022014, J074157,
J081329, J093844, J110742, J115644, J143455, and J172711. Of these,
J081329 and J172711 have Balmer-line profiles whose centroid is
somewhat blueshifted and with a strong blue peak and a weak red
shoulder. As such their classification is somewhat ambiguous.  The
remaining seven objects have very skewed Balmer-line profiles with one
clearly displaced peak, but a second peak is not clearly present,
making their classification as double-peaked emitters
questionable. Thus, we have retained all of these objects in our
sample. We note that J001224 was used as an illustrative example of
SBHB candidate by \cite{shen10a}.

\end{itemize}

In the next section we present the basic properties of the sample
objects and we study their spectra, paying particular attention to the
profiles of their H$\beta$ lines.

\section{Properties of the Sample}\label{S:sample-properties}

\subsection{SDSS Observations and Basic Properties of Sample
Objects}\label{S:basic-properties}

\begin{figure*}
\centerline{
\includegraphics[scale=1,angle=0]{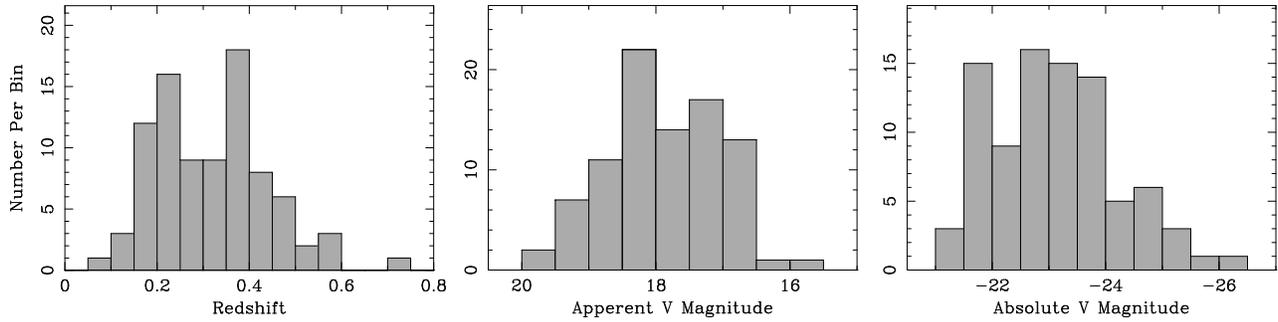}
}
\caption{Distribution of the basic properties of the objects in our
  sample.  The redshifts were determined from the wavelengths of the
  \flion{O}{3}{5007} line. The apparent $V$ magnitudes correspond to the
  PSF magnitude obtained from SDSS images. The absolute $V$ magnitudes
  take into account Galactic extinction towards the source. Further
  details are given in \S\ref{S:basic-properties} of the text.
\label{F:hist-properties}}
\end{figure*}

The apparent V magnitudes of the sample quasars were determined from
the SDSS PSF magnitudes in the $g$ and $r$ bands, using the
transformation equations of \citet{jester05}. We obtained the
redshifts of the sample quasars by measuring the peak wavelength of
the \flion{O}{3}{5007} line and by adopting a rest vacuum wavelength
of 5008.239\,\AA\ for this line (since the SDSS spectra are on a
vacuum wavelength scale). From this redshift, we determined the
distance modulus using the prescriptions of \cite{hogg99}\footnote{We
  used a program written by Benjamin Weiner, which is publically
  available at {\tt
    http://mingus.as.arizona.edu/$\sim$bjw/software/}. We compared the
  luminosity distances from this program to those of Ned Wright's
  cosmology calculator, available at {\tt
    http://www.astro.ucla.edu/\%7Ewright/CosmoCalc.html}, and found
  them to agree within 0.1\%.}, adopting the following cosmological
parameters: $H_0=73~{\rm km~s^{-1}~Mpc^{-1}}$, $\Omega_{\rm M}=0.27$,
and $\Omega_{\Lambda}=0.73$. Thus, we obtained the absolute $V$
magnitudes, taking into account Galactic extinction
\citep*[from][]{schlegel98}, but not $K$ corrections. The redshift,
apparent $V$ magnitude, and absolute $V$ magnitude distributions are
shown in the histograms of Figure~\ref{F:hist-properties}. We note
that the median redshift of our targets is 0.32, while the full range
of redshifts extends from 0.077 to 0.713.

In columns 6--10 of Table~\ref{T:obslog} we give a log of the
spectroscopic observations of the objects in our sample, including the
UT date of the observation, the telescope and instrument
configuration, the exposure time, the signal-to-noise ratio in the
spectrum ($S/N$), and the rest-frame wavelength range covered by the
spectrum. The table includes a separate block for each object, with
the first row of each block giving the particulars of the SDSS
observations. The instrument configurations are summarized in
Table~\ref{T:telcodes}. It is noteworthy that the spectral resolution
of the SDSS spectra varies with wavelength such that the velocity
resolution improves linearly with the log of the wavelength \citep[see
  the discussion and illustration in Appendix B and Figure~16
  of][]{bernardi03}. Thus, in Table~\ref{T:telcodes} we list the
spectral resolution at 6400\,\AA, which corresponds to the location of
the H$\beta$ line for the median redshift of our sample. The $S/N$ was
determined from the fluctuations about the mean continuum level in a
50\,\AA\ wide window near the emission line of interest.  In
particular, when the spectrum includes the H$\beta$ line we give the
$S/N$ in the continuum near this line, at 4600\,\AA\ ($\Delta v \approx
-16,600\kms$ relative to H$\beta$).  If the spectrum includes only the
\ion{Mg}{2}~$\lambda$2800 line, we give the $S/N$ in the continuum
near this line, at 2900\,\AA\ ($\Delta v \approx
+10,500$\kms\ relative to \ion{Mg}{2}).

\begin{figure*}
\centerline{
\includegraphics[scale=0.7,angle=0]{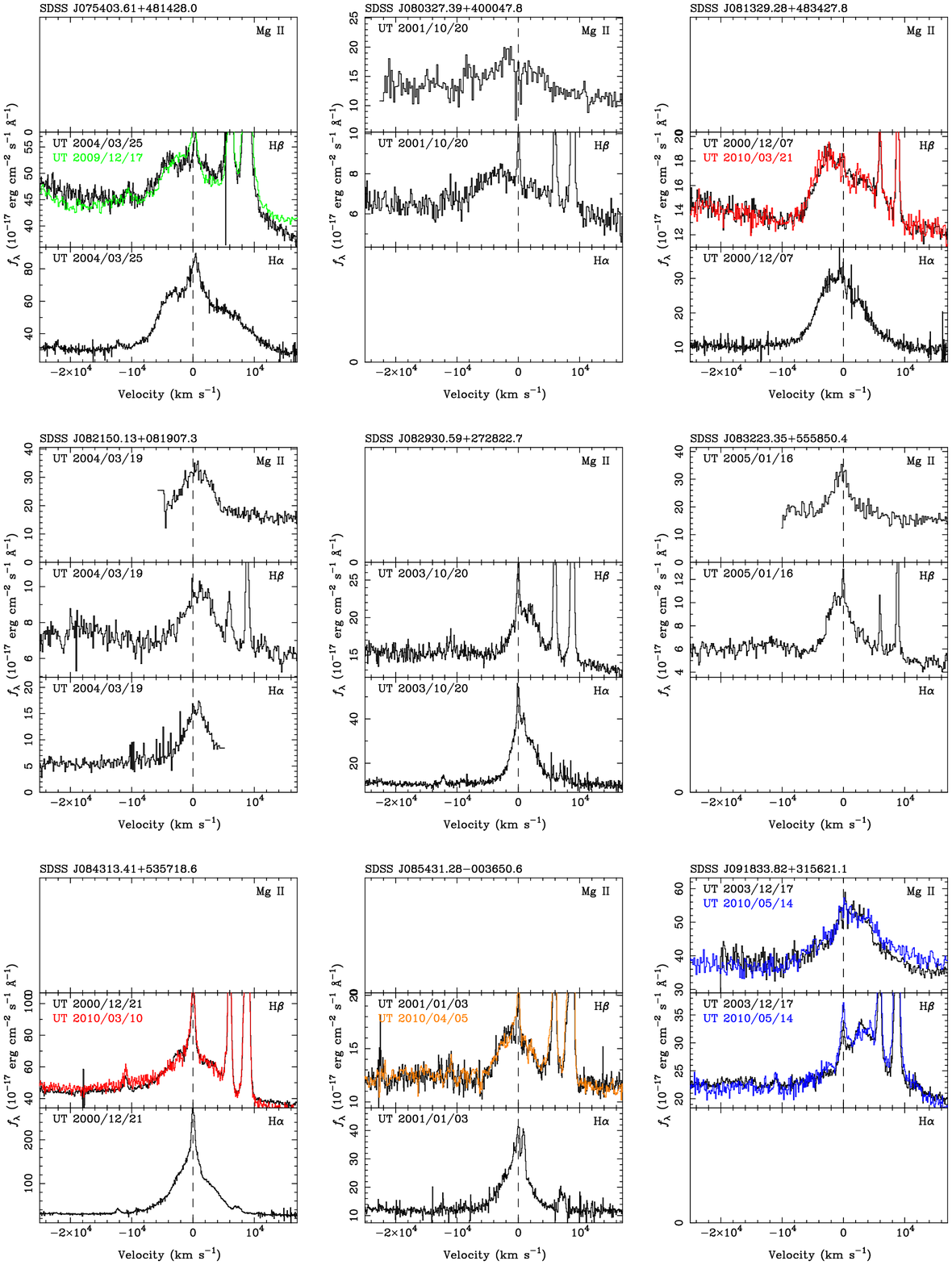}
}
\caption{{\bf ABRIDGED} \ion{Mg}{2}, H$\beta$, and H$\alpha$ spectra
  of our target objects. For each object we show a 3-panel figure with
  the profiles of all three lines on a common velocity scale. The
  original SDSS spectra are plotted in black, while the post-SDSS
  spectra are color-coded as follows: KPNO 4m spectra are in red,
  Palomar 5m spectra are in blue, MDM 2.4m spectra are in green, and
  HET spectra are are in orange. For the purpose of this illustration,
  some spectra were binned to increase the $S/N$ per bin. The vertical
  scale was adjusted specifically for the original SDSS
  spectra. Post-SDSS spectra have been scaled and shifted so that the
  profiles of the broad lines match those of the SDSS spectra as
  closely as possible; see equation (\ref{Q:linear}) and
  \S\ref{S:comparison} of the text. As a result of this scaling, there
  is an apparent difference in the strengths of the narrow lines, if
  the flux of the broad line has varied between epochs.
\label{F:spectra}}
\end{figure*}

The spectra of the \ion{Mg}{2}, H$\beta$, and H$\alpha$ lines of the
objects in our sample are shown in Figure~\ref{F:spectra} on a common
velocity scale. The broad-line spectra shown in this figure illustrate
that it is easier to find broad displaced peaks in the H$\beta$
profiles, especially when the displacement is small, because the peak
of the broad H$\alpha$ profile is ``contaminated'' by the narrow
H$\alpha$+\fion{N}{2} complex. For example, in J130534 and J140007,
one can easily discern an offset broad peak in the H$\beta$ but not in
the H$\alpha$ profiles. Vice versa, the H$\alpha$ profiles show more
clearly the shape of the red side of the profile where the H$\beta$
line may suffer from severe contamination by the \fion{O}{3} doublet.
For example, in J075403, J091928, and J121113, what appears to be an
extended red wing in the H$\beta$ profile actually has the form of a
shoulder or weak red peak in the H$\alpha$ profile The profiles of the
\ion{Mg}{2} lines are very similar to the profiles of the Balmer
lines. Specifically, there is no discernible misalignment of the peaks
of the different lines with only one exception: J105203.  One must
keep in mind, however, that the contamination of the peak of the broad
line by the narrow doublet is more severe than in the case of H$\beta$
and comparable in severity to the case of H$\alpha$.  Moreover, since
the \ion{Mg}{2} lines are resonance lines, the profiles of the broad
emission lines are often contaminated by associated {\it absorption}
lines, which can be fairly strong.

\subsection{Processing of SDSS Spectra and Emission-Line 
Measurements}\label{S:sdss-lines}

To quantify the spectroscopic properties of our targets, we measured
the properties of some of the strong narrow emission lines as well as
the properties of the profiles of the broad H$\beta$ lines from the
SDSS spectra. 

Specifically, we measured the integrated fluxes and full widths at
half maximum (FWHM) of the following narrow lines:
\flion{Ne}{5}{3426}, \flion{O}{2}{3726,3729}, \flion{Ne}{3}{3869},
H$\beta$ and \flion{O}{3}{5007}. The widths and relative strengths of
the narrow lines, especially the high-ionization ones, can serve as
tests of the rapidly recoiling BH hypothesis, as discussed, for
example, by \citet{bonning07}, \citet{komossa08a} and
\citet{shields09a}. To isolate the lines, we fitted a low order
polynomial to the local {\it effective} continuum. In the case of the
H$\beta$ and \flion{O}{3}{5007}, we used a continuum model that also
includes the contribution of the \ion{Fe}{2} lines, as we describe
later in this section. The line fluxes were then determined by
integrating the observed line profile, while the widths were
determined by fitting a Gaussian to the line profile and then
correcting the FWHM for the finite resolution of the spectrograph at
the observed wavelength of the line \citep[see Appendix B and
  Figure~16 of][]{bernardi03}. We made no effort to de-blend the two
lines in the \fllion{O}{2}{3726,3729} doublet, therefore we do not
report their widths. However, we did measure the width of the blended
doublet, which allows us to discern in some cases whether the
\fion{O}{2} lines are narrower than other forbidden lines. The
observed line fluxes were corrected for Galactic extinction using the
values of $A_{\rm V}$ listed in Table~\ref{T:obslog} and assuming the
extinction law of \citet{seaton79}. The resulting measurements are
reported in Table~\ref{T:narrow} where we give the line fluxes
relative to that of the \flion{O}{3}{5007} line before and after
extinction corrections, the observed and corrected \flion{O}{3}{5007}
flux, and the corrected \flion{O}{3}{5007} luminosity, as well as the
FWHM of the lines.

\begin{figure}
\centerline{
\includegraphics[scale=0.5,angle=0]{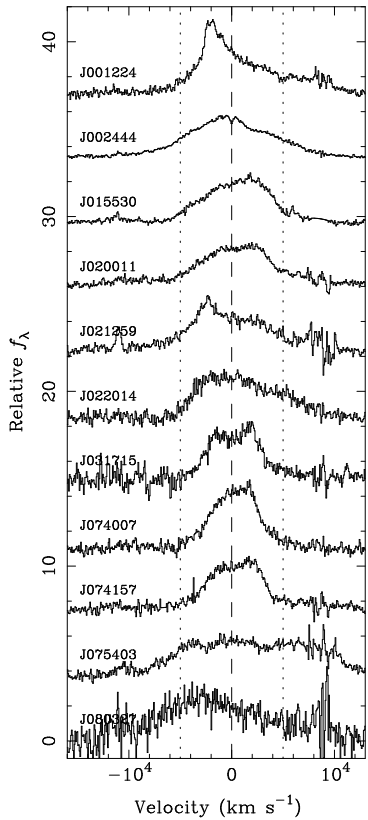}
}
\caption{{\bf ABRIDGED} Profiles of the broad H$\beta$ lines from the
  SDSS spectra, shown on a common velocity scale. The narrow H$\beta$
  and \fion{O}{3} lines and underlying continuum were subtracted as
  described in \S\ref{S:sdss-lines} of the text. The resulting
  profiles were normalized arbitrarily and offset vertically from each
  other for clarity. The vertical dashed line shows the location of
  the narrow H$\beta$ line, while the vertical dotted lines identify a
  window of $\pm 5,000$\kms\ from this line. The noise often seen at
  +8875\kms\ is a result of imperfect subtraction of the
  \flion{O}{3}{5007} line. The \lion{He}{2}{4686} line, which was not
  subtracted, is sometimes discernible at $-10,800$\kms.
\label{F:stack}}
\end{figure}

In order to isolate the profiles of the broad H$\beta$ lines we first
subtracted a model of the underlying continuum and \ion{Fe}{2}
multiplets, which was fitted to the spectral range
4,000--5700~\AA. The underlying continuum consisted of a linear
combination of a starlight template and a featureless power law while
the \ion{Fe}{2} multiplets were modeled by suitably broadening a
template, as described in \citet{boroson92}. As a last step we
subtracted the narrow H$\beta$ and \fion{O}{3} lines by using the
profile of the \flion{O}{3}{5007} line as a template for the other two
lines. As a cross-check, we used the profile of the \flion{O}{3}{4959}
as a template to subtract the \flion{O}{3}{5007} line. The resulting
broad H$\beta$ profiles are shown on a common velocity scale in
Figure~\ref{F:stack}.

After isolating the broad H$\beta$ line profiles we quantified their
shapes by measuring some of their central moments and related
quantities. The $n^{\rm th}$ {\it central} moment of a line profile,
$\mu_n$, is defined in terms of the wavelengths and flux densities of
the discrete pixels in the line profiles ($\lambda_i$ and $f_i$
respectively) as $ \mu_n \equiv K \sum \left(\lambda_i -
\langle\lambda\rangle\right)^n f_i $, where $K$ is a normalization
constant defined by $1/K = \sum f_i$ and $\langle\lambda\rangle \equiv
K \sum \lambda_i \, f_i$ is the first moment or {\it centroid} of the
line profile.  Using the second and third central moments we
determined the standard deviation and skewness of the line profiles as
$\sigma = \mu_2^{1/2}$ and $s = \mu_3/\mu_2^{3/2}$, respectively. We
also determined the Pearson skewness coefficient, defined as $p =
(\langle\lambda\rangle - \lambda_m)/\sigma$, where $\lambda_m$ is the
median wavelength\footnote{Because of the definitions we adopted for
  $s$ and $p$, $s\propto -p$.}. In Table~\ref{T:moments} we report
the centroid velocity shift relative to the nominal wavelength of
H$\beta$ using the relativistic Doppler formula, the velocity
dispersion of the line, computed from the standard deviation as
$c\sigma/\lambda_0$ (where $\lambda_0$ is the nominal wavelength of
H$\beta$ and $c$ is the speed of light), the skewness, and the Pearson
skewness coefficient.

\input  tab3.tex
\input  tab4.tex

In addition to the above quantities, we also measured the velocity
offset of the peak of the broad line and its FWHM, which we also
report in Table~\ref{T:moments}. In order to make these measurements,
we had to locate the peak of the broad line, which is somewhat
subjective since the observed profiles have a finite $S/N$.  Thus we
attempted to determine the location of the peak of the broad line
profile by fitting it with a Gaussian. This procedure leads to an
uncertainty because the symmetric Gaussian does not always fit the
line peak well, especially when the peak is asymmetric. To quantify
this uncertainty, we made several measurements of the same line,
varying the region around the peak used for the fit. In the end we
adopted the maximum and minimum peak wavelengths from our trials to
define the range of possible values and took their average as the best
estimate of the peak wavelength. Thus, in Table~\ref{T:moments} we
report the velocity offset of the peak, computed from the relativistic
Doppler formula, as well as its uncertainty. Finally, the FWHM was
determined based on the height of the peak and expressed as a velocity
width in the same manner.

\subsection{Demographics and Statistical Properties of the Broad 
H$\beta$ Profiles and the Narrow Lines}\label{S:demographics}
 
\begin{figure}
\centerline{\includegraphics[scale=0.45,angle=0]{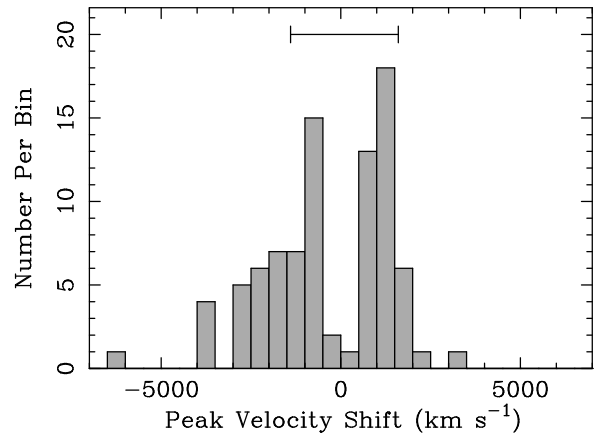}}
\caption{Distribution of the velocity shifts of the peaks of the broad
  H$\beta$, measured as described in \S\ref{S:sdss-lines} of the text.
  The dearth of objects in the rage $\pm 500$\kms\ is a selection
  effect resulting from our inability to detect such small shifts with
  our method. The horizontal error bar at the top of the frame
  indicates the range of shifts found by \citet{bonning07}: $\pm 3$
  standard deviations about a mean of 100\kms.
\label{F:hist-shift}}
\end{figure}

Figure~\ref{F:hist-shift} illustrates the distribution of shifts of
the peaks of the broad H$\beta$ lines among objects in our sample. The
dearth of objects in the range $\pm 500$\kms\ is a result of the fact
that we purposely excluded most objects with such small shifts.  In
the same figure we show the range of shifts ($\pm3$ standard
deviations about a mean of 100\kms) measured by \citet{bonning07}, who
studied the H$\beta$ profiles of 2598 quasars at $0.1<z<0.81$ drawn
from a sample of approximately 13,000 objects from the SDSS DR5 and
found 9 cases of large offsets, from $\approx 1,000$\kms to a maximum
of 2,667\kms.  Our study appears to complement and supplement theirs
since we recover a substantially larger number of highly offset peaks:
32 objects with $|\Delta v|>1500\kms$, a factor of 8 larger than what
we would predict just by scaling the sample sizes. Similar conclusions
follow if we compare our results with those of \citet{zamfir10}, who
studied the high-$S/N$ spectra of 469 quasars with $z < 0.7$ from the
SDSS DR5. The vast majority of our objects belong to their population
B \citep[FWHM$_{\rm H\beta}>4,000\kms$, weak \ion{Fe}{2} lines;
  see][]{sulentic02}, whose distribution of broad H$\beta$ shifts
shows some preference for redshifts and has extrema of
$\pm3100\kms$. Thus, the objects we have identified here occupy the
wings of the distribution of shifts of the peak of the broad H$\beta$
line.

Perusal of Figures~\ref{F:spectra} and \ref{F:stack} leads us to
classify the Balmer line profiles into three broad families according
to their shapes. The first family includes profiles whose widths are
smaller than average and visual inspection does not show them to be
particularly asymmetric.  Two of the objects discussed extensively in
the recent literature, J092712 and J105041, are prime examples of
objects in this family. Additional examples include J082930, J095036,
J110050, J122811, 132704, and J162914. The second family includes
skewed profiles showing an extended wing or a shoulder on the side
opposite from the direction of their displacement (e.g., profiles with
a blue-shifted peak with an extended red wing or a weak red shoulder).
Representative examples of this family with blue-shifted peaks (and
red asymmetries) are J001224, J091928, J115449, J123001, J125809,
J133432, J152942, J153636, while examples with red-shifted peaks (and
blue asymmetries) include J015530, J093653, J094603, J113651, J130534,
J143455. The third family includes objects with flat-topped profiles,
such as J020011, J031715, J074157, J095539, J112751, J154340, J163020.

\begin{figure}
\centerline{\includegraphics[scale=0.4,angle=0]{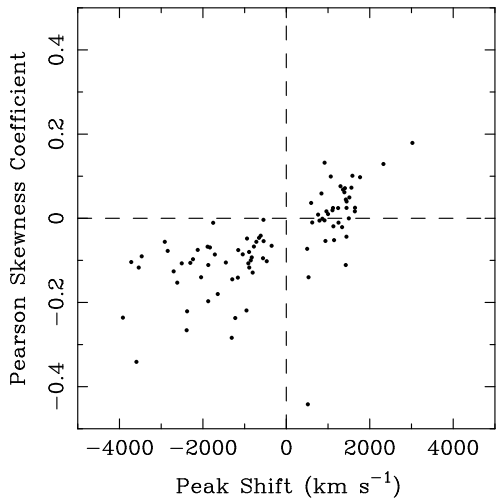}}
\caption{Relation between the skewness of the broad H$\beta$ profiles
  and the shift of their peaks. The Pearson skewness coefficient and
  the shift of the peak were measured as described in
  \S\ref{S:sdss-lines} and are tabulated in Table~\ref{T:moments}.
\label{F:shift-skew}}
\end{figure}

In addition to the visual classification of line profiles we have also
looked for trends in their quantitative properties and we found a
correlation between their shift and their skewness. We illustrate this
correlation in Figure~\ref{F:shift-skew}, where we plot the Pearson
skewness coefficient {\it vs} the peak shift.  A similar correlation
is recovered when we consider other pairs of quantities that track
skewness and shift, such as the skewness coefficient and the centroid
shift. As the shift of the lines increases, the profile becomes more
skewed. Moreover, the asymmetry manifests itself as an extended wing
or weak shoulder on the side of the line profile that is {\it
  opposite} from the direction of the shift, with the result that the
sign of the skewness tracks the sign of the shift. This correlation
quantifies the trend we noted in the second family of line profiles in
the previous paragraph. We have not found any correlation between the
FWHM of the lines and the shift of their peaks nor between any other
pairs of quantities that track width and shift. \citet{bonning07} do
report that H$\beta$ profiles with large shifts also have large FWHM
but the vast majority of objects in their sample show considerably
smaller H$\beta$ shifts than the objects in our sample and the shifts
they regard as large are considerably smaller than the shifts we
measure. Therefore, our results are not at odds with theirs.  We did
find a weak correlation between the width and the skewness. These
correlations are consistent with the findings of \citet{zamfir10},
shown in their Figures 6a and 6c.

\begin{figure*}
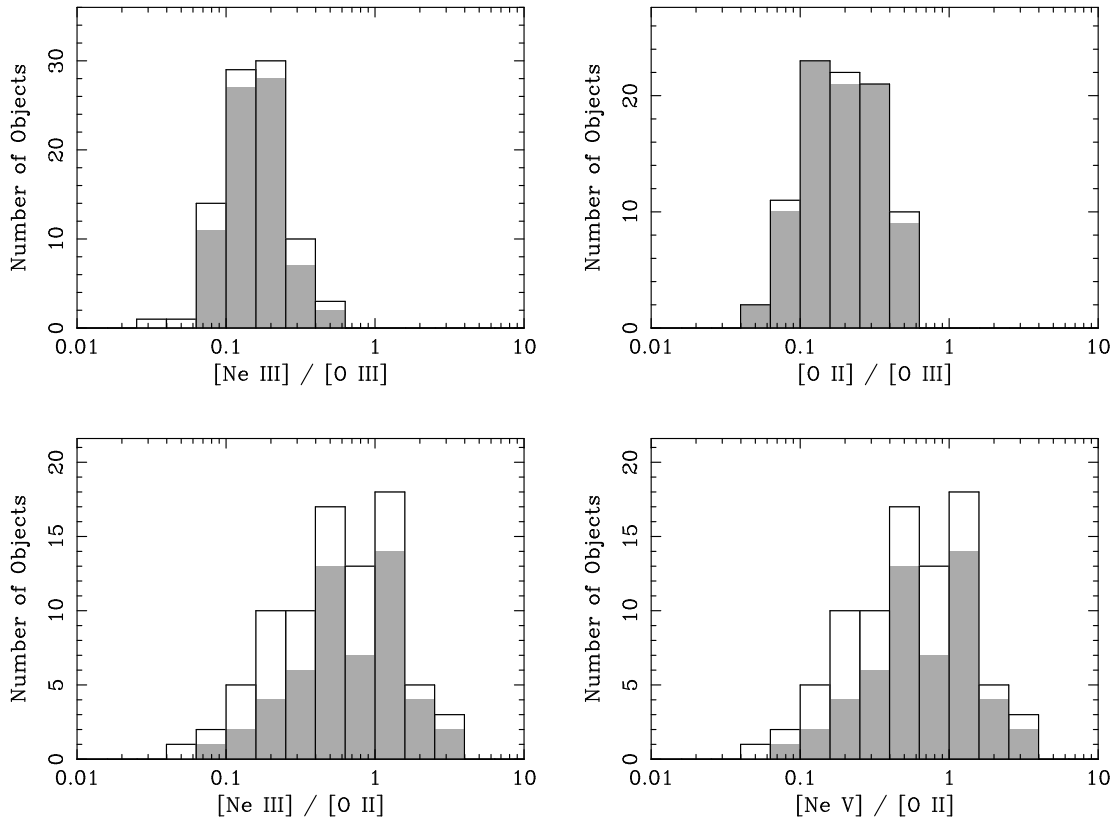

\centerline{\includegraphics[scale=0.4,angle=0]{f7a.eps}\hskip 0.25truein
            \includegraphics[scale=0.4,angle=0]{f7b.eps}}\vskip 0.25truein
\centerline{\includegraphics[scale=0.4,angle=0]{f7c.eps}\hskip 0.25truein
            \includegraphics[scale=0.4,angle=0]{f7d.eps}}
\caption{Distribution of four different narrow-line ratios among the
  quasars in our sample, after correction for Galactic extinction.
  The measurements were made from the SDSS spectra, as described in
  \S\ref{S:sdss-lines} of the text, and the results are tabulated in
  Table~\ref{T:narrow}. The shaded portions of bins represent actual
  measurements while the hollow portions represent limits.
\label{F:narrow-ratios}}
\end{figure*}

\begin{figure}
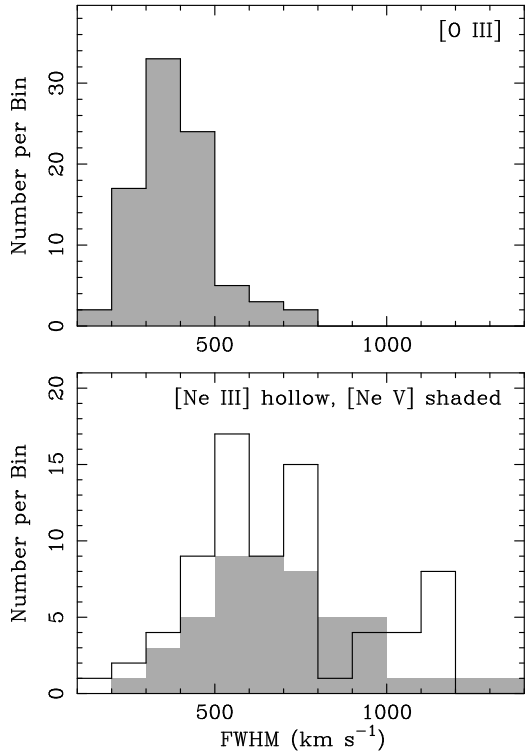

\centerline{\includegraphics[scale=0.4,angle=0]{f8a.eps}}\medskip
\centerline{\includegraphics[scale=0.4,angle=0]{f8b.eps}}
\caption{Distribution of the FWHM of high-excitation narrow lines of
  quasars in our sample, after correction for the finite resolution of
  the spectrograph.  The measurements were made from the SDSS spectra,
  as described in \S\ref{S:sdss-lines} of the text, and the results
  are tabulated in Table~\ref{T:narrow}. The upper panel shows the
  distribution of widths of the \flion{O}{3}{5007} lines, while the
  lower panel shows the distributions of \flion{Ne}{3}{3869} (black,
  solid line with hollow bins) and \flion{Ne}{5}{3426} (shaded bins
  with no outline).
\label{F:narrow-widths}}
\end{figure}

In Figures~\ref{F:narrow-ratios} and \ref{F:narrow-widths} we present
the distribution of relative intensities and FWHM of the the narrow
lines using the measurements presented in Table~\ref{T:narrow}. The
relative intensity distributions resemble very closely those of
Seyfert~1 and Seyfert~1.5 galaxies in the compilation of
\citet{nagao01}, which suggests that the ionization parameter and the
shape of the ionizing continuum are similar to those of Seyfert
galaxies. The distribution of FWHM of the narrow lines is similar to
what is observed in Seyfert galaxies \citep*[see][]{whittle85,
  moore96} and is also evident in the composite quasar spectrum of
\citet{vandenberk01}.  The average FWHM of the \fion{O}{3} line is
very similar to that found by \citet{salviander07} in their study of
the spectra of 1736 quasars from SDSS DR3.  Both the \fion{Ne}{5} and
\fion{Ne}{3} lines in our sample are broader than \fion{O}{3}, on
average, and the \fion{Ne}{5} lines are always as broad as or broader
than the \fion{O}{3} line within errors.  This trend follows the
correlation between FWHM and critical density noted for Syeferts
\citep{filippenko84,whittle85}. Thus the objects in our collection do
not appear to stand out from typical active galactic nuclei as far as
the properties of the narrow lines are concerned.

\section{Followup Observations and New Spectra}\label{S:newobs}

\subsection{Data Acquisition and Reduction}\label{S:reduction}

We carried out followup observations of 68 objects from our sample
using four different telescopes: the brightest objects were observed
with the 2.4m Hiltner telescope at MDM observatory, objects of
intermediate brightness were observed with the 4m Mayall telescope at
Kitt Peak National Observatory and the 5m Hale telescope at Palomar
Observatory, while the faintest objects were observed with the 9.2m
Hobby-Eberly Telescope (HET). With the followup observations we
specifically targeted the H$\beta$ lines of the target objects,
although we also did attempt to obtain spectra of the H$\alpha$ and
\ion{Mg}{2} lines whenever possible.  The spectrographs and
configurations we used for the observations are summarized in
Table~\ref{T:telcodes}, along with the spectral resolution that each
configuration can attain. The log of observations is included in
Table~\ref{T:obslog} where, in addition to the observation date and
exposure time, we also report the rest-frame wavelength range covered
by each spectrum and the $S/N$ achieved in the continuum near the line
of interest (measured as described in \S\ref{S:selection}, above). The
weather conditions were varied during the followup observations and
some spectra were taken through clouds.  As a result, there are
instances where we obtained a lower $S/N$ than the original SDSS
spectrum of the same objects in a comparable exposure time.

The spectra were reduced in a standard manner using
IRAF\footnote{IRAF, the Interactive Reduction and Analysis Facility,
  is distributed by the National Optical Astronomy Observatory, which
  is operated by the Association of Universities for Research in
  Astronomy (AURA) under cooperative agreement with the National
  Science Foundation (see {\tt http://iraf.noao.edu/}).} for the first
steps of the reduction process (bias subtraction, flat field division,
sky subtraction, and extraction of raw spectra) and our own programs
for the subsequent calibration (wavelength calibration, correction for
continuous atmospheric extinction and discrete absorption bands, and
flux calibration). Accurate wavelength calibration is important for
our purposes since we aim to look for shifts in the H$\beta$ lines
between the SDSS and post-SDSS spectra; thus we give here the details
of the process. To obtain the transformation between pixel number and
observed (air) wavelength we fitted a polynomial to the table of arc
line pixel location {\it vs} wavelength. Depending on the instrument,
we used 20--65 arc lines and achieved a fit with root-mean-square
(r.m.s.)  residuals of better than 0.10 pixels with a polynomial of
order 4 or lower. Thus, the relative wavelength scale is good to
10\,\kms\ for the KPNO 4m and Palomar 5m spectra, 13\,\kms\ for the
HET spectra and 18\,\kms\ for the MDM 2.4m spectra. In comparison, the
relative wavelength scale of the SDSS spectra is good to 0.07 pixels
or 5\,\kms\ (r.m.s.). After the spectra were fully calibrated, we
applied heliocentric corrections (amounting to 0.15\,\AA\ or less) and
converted the wavelength scale from air wavelengths to vacuum
wavelengths, following the convention for SDSS spectra. For the last
step we used the relation between vacuum and air wavelengths given by
\citet{morton91}, which yields corrections between approximately 0.5
and 2.5\,\AA\ over the wavelength range of our spectra. With these
corrections the r.m.s. shift between the SDSS and post-SDSS spectra
amounted to 0.67\,\AA\ or 41\,\kms, determined by cross-correlation of
the spectra over the region around the \fllion{O}{3}{4959,5007}
lines. Thus, in a final step we refined the alignment of spectra by
rectifying these residual shifts. The final alignment is limited by
our ability to determine the shift of the \fion{O}{3} lines by
cross-correlation. In about 80\% of the cases the r.m.s. alignment
uncertainty is 7\kms\ or better, while in the other 20\% of cases it
lies between 7 and 25\kms.

\subsection{Qualitative Comparison of New and Old Spectra}\label{S:comparison}

\begin{figure}
\centerline{\includegraphics[scale=0.4,angle=0]{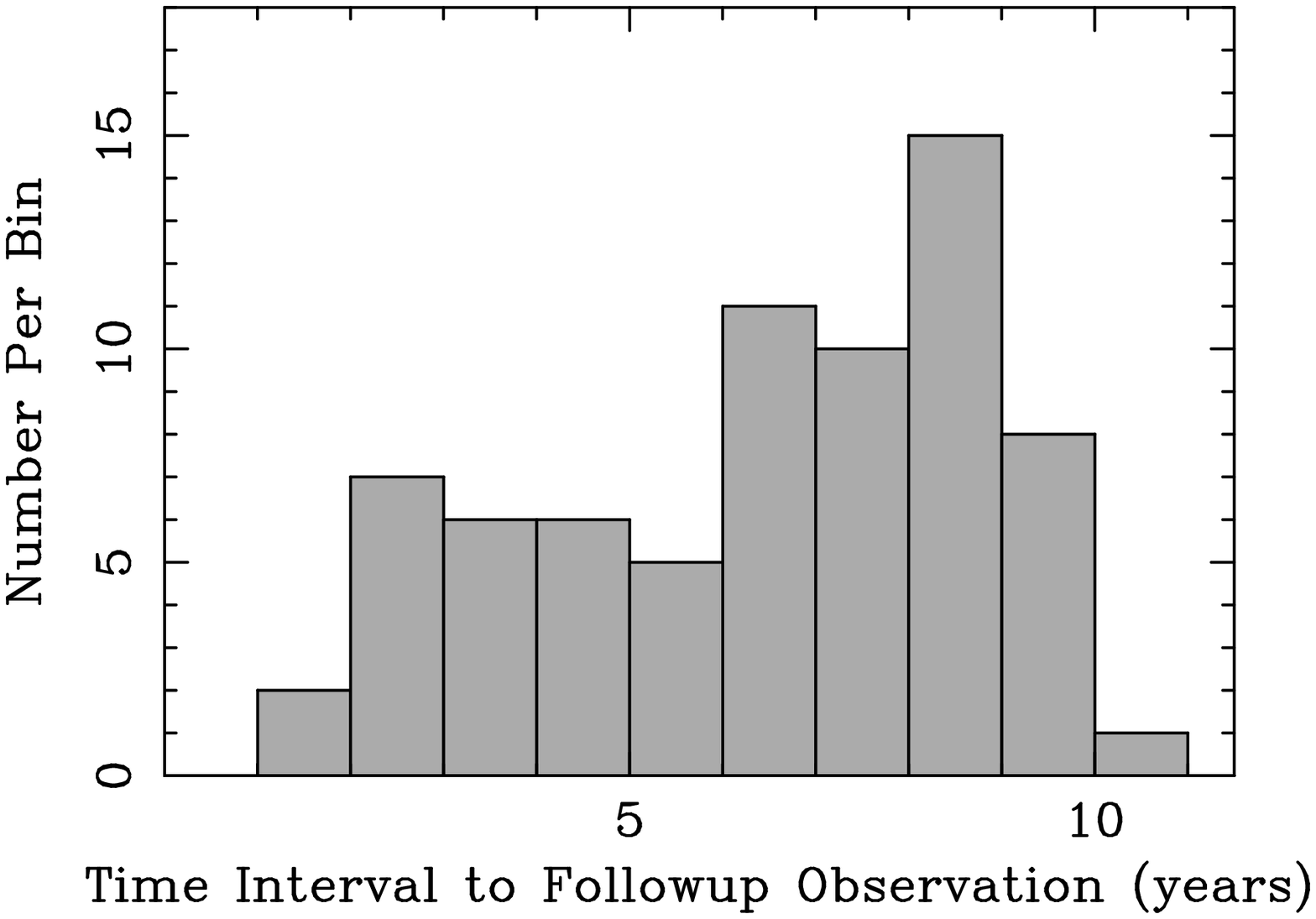}}
\caption{Distribution of the time intervals \redtext{(in the
    observer's frame)} between the original SDSS spectra and the
  followup spectra obtained here. The median of this distribution is
  7.0 years. For 70\% of the objects the time interval between
  observations is at least 5.7 years.
\label{F:hist-intervals}}
\end{figure}

The distribution of time intervals \redtext{(in the observer's frame)}
between the original SDSS observations and the followup observations
is shown in Figure~\ref{F:hist-intervals}. The median of the time
interval distribution is 7.0 years, with 70\% of the objects having a
time interval between observations of at least 5.7 years.

In Figure~\ref{F:spectra} we overplot the post-SDSS spectra on the
original SDSS spectra for comparison. Since our primary goal is to
look for changes in the profiles of the broad emission lines, we have
scaled and shifted the post-SDSS spectra so as to match these profiles
as closely as possible. We applied a transformation to the flux
density scale of the form
\begin{equation}
f_\lambda^{\prime} = af_\lambda + b_\lambda \; ,
\label{Q:linear}
\end{equation}
where $a$ and $b_\lambda$ are constants, determined by minimizing the
differences between the SDSS and post-SDSS spectra in selected regions
of the broad line profiles and the adjacent continuum. This scaling
convention highlights changes in the profiles of the two spectra but
it hides changes in the flux of the broad line and the
continuum. Changes in the flux of the broad line can still be
discerned by examining the intensities of the narrow lines (most
notably \fion{O}{3} ), which are not expected to change in the time
interval between observations. In other words an apparent mismatch in
the strength of the narrow lines between two spectra indicates that
flux of the broad line has changed between observations.

An inspection of Figure~\ref{F:spectra} reveals changes in the broad
H$\beta$ lines between the SDSS and post-SDSS observations in many
cases. Changes in the profile shapes are the easiest to
discern. Examples of objects displaying this type of variability
include J093653, J112751, J113706, J140700, and J143123.  There are
also several cases of changes in the integrated flux of the line,
discernible through mismatches in the strengths of the \fion{O}{3}
lines, including but not limited to J021259, J094603, J111329,
J115449, J132704, and J143455. Yet another category of changes
comprises cases where the broad line has shifted in
wavelength/velocity by a small amount. The most obvious example of
such a change is J095036, while J094603, J140251, and J180545 represent
examples of more subtle shifts of this type. This type of variability
if, of course, of particular interest here, therefore we study it
further in \S\ref{S:shifts}, below.

\section{Search for Velocity Changes in the Broad H$\beta$ Profiles}\label{S:shifts}

\subsection{A $\chi^2$ Cross-Correlation Method for Determining 
Velocity Changes}\label{S:shifts:method}

To determine any changes in the velocities the offset peaks of the
broad H$\beta$ lines between the SDSS and post-SDSS spectra we
developed an algorithm that is a variant of the standard
cross-correlation method. In our method we shift one of the two
spectra in small steps and at each step we compare it with the other
spectrum via the $\chi^2$ test. Based on several tests, we prefer this
method over the standard cross-correlation (which employs the overlap
integral rather than the $\chi^2$ test at each step) for the following
reasons: (a) it allows us to check whether the profile of the H$\beta$
line has changed between epochs (this is important because small
changes in the profile can mimic a velocity change), (b) the $\chi^2$
minimum is much sharper and deeper than the overlap integral maximum,
thus easier to locate, especially if we are relying on the wings of
the broad line for our comparison, (c) it allows us to determine
uncertainties and limits on the shift in a more straightforward and
less computationally intensive way, without the need for extensive
simulations for each pair of spectra being compared.

\begin{figure*}
\leftline{
\includegraphics[scale=0.33,angle=0]{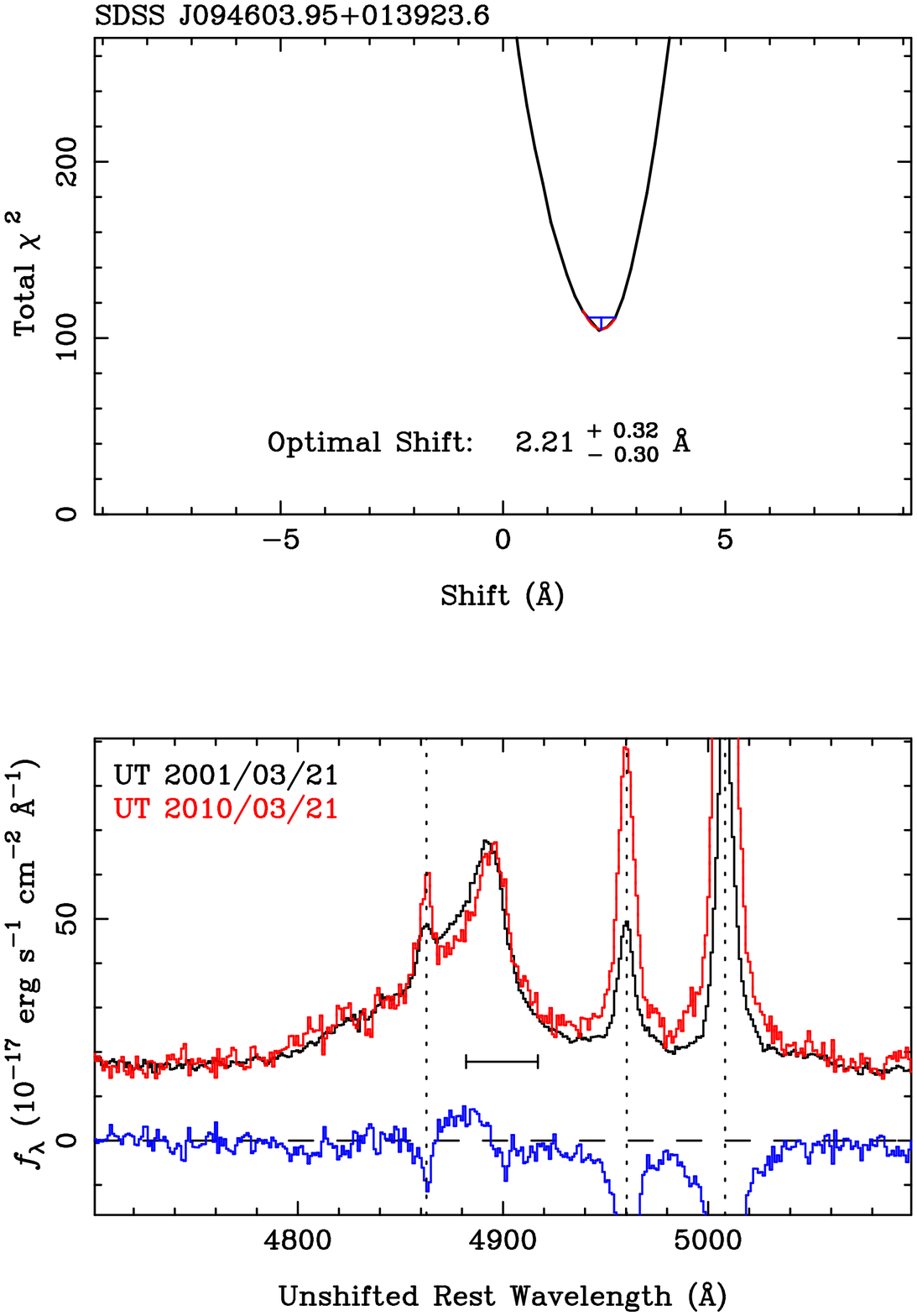}\hskip 0.12truein
\includegraphics[scale=0.33,angle=0]{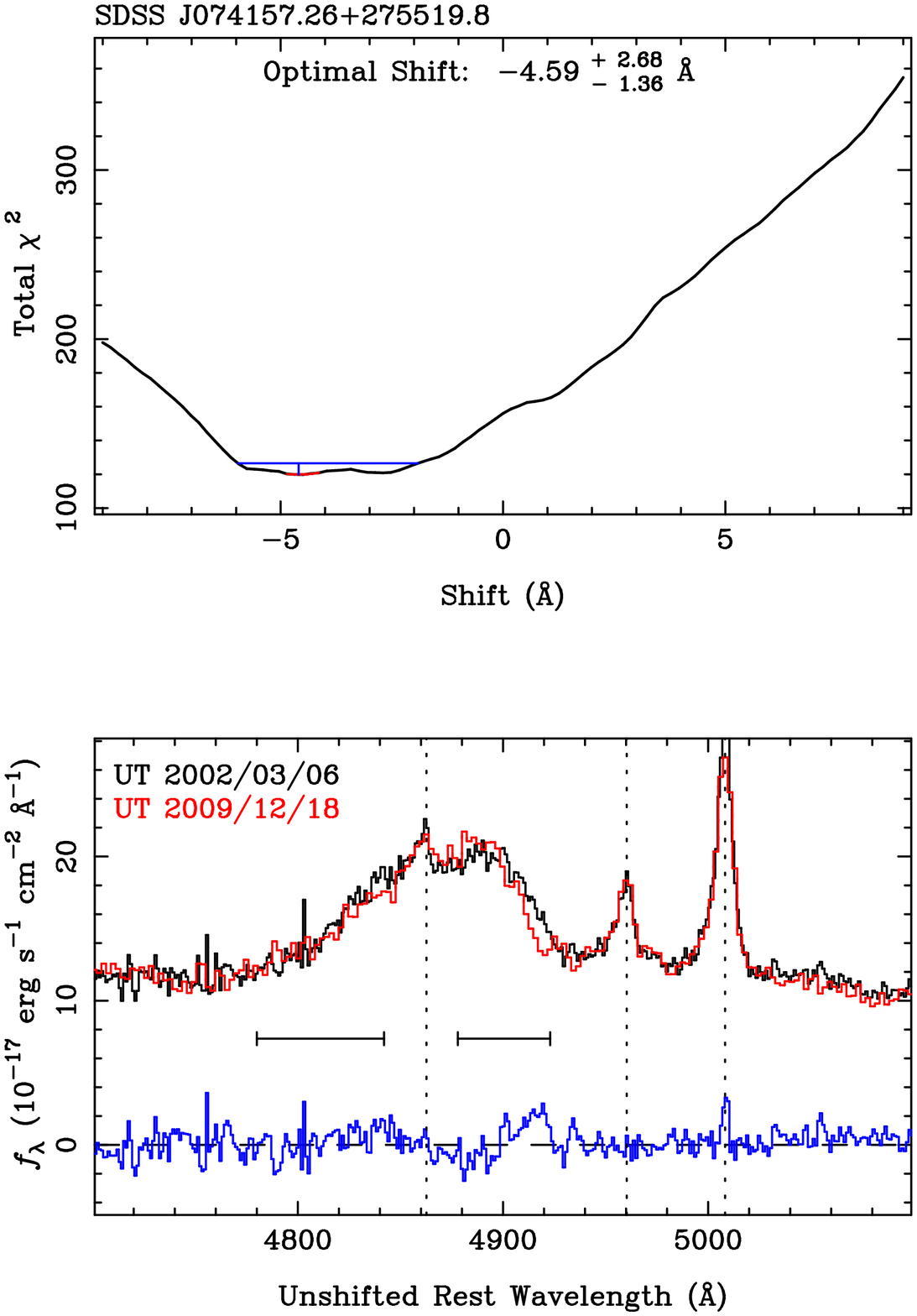}\hskip 0.12truein
\includegraphics[scale=0.33,angle=0]{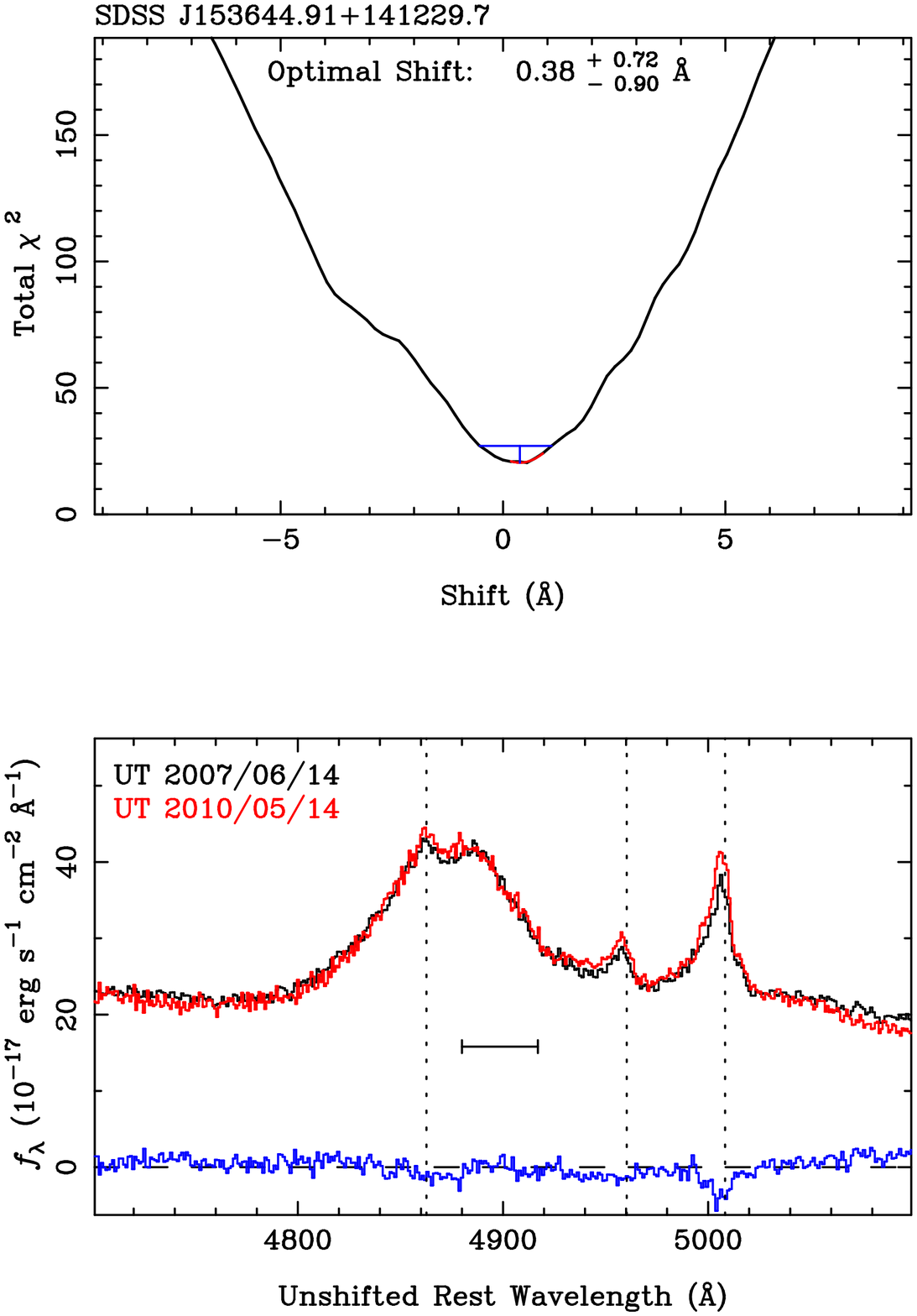}
}
\caption{
  Examples of the
  application of our cross-correlation technique (see
  \S\ref{S:shifts:method} of the text) to three pairs of spectra. In
  the lower panel of each set we show the the two spectra
  appropriately scaled and superposed for comparison. The original
  SDSS spectrum is shown in black, the post-SDSS spectrum (which is
  the one that is scaled, shifted and rebinned by the
  cross-correlation algorithm) is shown in red, while the residuals
  after subtraction of the two is shown in blue (in this illustration
  the residuals represent the difference between the {\it un-shifted}
  spectra). Since the scaling of the two spectra is such that the
  continuum and broad lines match, there is typically a mismatch
  between the narrow lines, which indicates that the continuum and/or
  the broad lines have varied between the two epochs. The vertical
  dotted lines mark the nominal positions of the narrow H$\beta$ and
  \fion{O}{3} lines. The horizontal error bars show the sections of
  the broad line profiles used by the cross-correlation algorithm. The
  upper panel of each set shows the variation of $\chi^2$ with the
  shift applied to the post-SDSS (red) spectrum. The red arc at the
  trough of the $\chi^2$ curve shows the global minimum and the (blue)
  horizontal line shows the 99\% confidence level for one interesting
  parameter, $\chi^2_{\rm min}+6.63$. We note for reference that a
  wavelength shift of 1~\AA\ corresponds to a velocity shift of
  62\kms. The three examples shown here illustrate the following: {\it
    Left:} a significant shift of the offset peak of the broad line;
  {\it Middle:} a subtle profile variation that mimics a shift of the
  broad line in the $\chi^2$ curve; {\it Right:} no significant shift
  of the offset peak.
\label{F:crosscorr}}
\end{figure*}

Our $\chi^2$ cross-correlation method is applied to a pair of spectra
as follows: before any comparison is made, a linear transformation is
applied to the flux scale of the second spectrum of the pair following
equation~(\ref{Q:linear}), in order to match the continuum and the
broad H$\beta$ line. This transformation creates a mismatch in the
narrow lines, of course, but it allows us to carry out a direct
comparison of the broad lines, without subtracting the continuum or
the narrow lines, which would introduce additional systematic
errors. The wavelength scale of the second spectrum is then shifted in
small steps (typically 0.1--0.2~\AA), rebinned to the wavelength scale
of the first spectrum, and finally the $\chi^2$ statistic is evaluated
to compare the profiles of the broad H$\beta$ lines in one or two
windows that are not affected by narrow lines. These windows are
chosen to encompass the offset broad peak and/or the steep wings or
shoulders of the line so as to maximize the sensitivity to small
shifts. The final output is a $\chi^2$ curve as a function of
shift. The minimum of this curve, determined by fitting a parabola to
the five lowest points, represents the optimal wavelength shift. The
99\% (or ``$2.6\,\sigma$'') confidence interval about the minimum (for
one interesting parameter) is determined by finding the shifts that
correspond to $\chi^2_{\rm min}+6.63$ \citep*[see][]{lampton76}.  In
Figure~\ref{F:crosscorr} we show graphically the application of this
technique in three example cases, a shift that is statistically
significant, a shift that is not statistically significant, and an
apparent shift that is caused by a variation in the broad H$\beta$
profile between the two epochs.

We verified through simulations that $\chi^2_{\rm min}+6.63$ does
indeed reflect the 99\% confidence interval.  Each simulation was
carried out by creating a copy of an observed high-$S/N$ spectrum,
adding synthetic noise, and then measuring the shift between the two
spectra. We repeated this test for several examples of observed
spectra from our collection.  In 1000 realizations in each case we
found that about 1\% of the optimal shifts departed from zero by more
than the amount predicted by the $\chi^2_{\rm min}+6.63$ condition, as
expected. An additional check of the procedure was carried out by
switching the order of the spectra supplied to the $\chi^2$
cross-correlation algorithm and comparing the resulting shifts. Indeed
the distribution of discrepancies between the two shifts, normalized
by the 99\% confidence error bar was found to be approximately
Gaussian with a standard deviation of 1/2.6, as expected.

\subsection{Application of the $\chi^2$ Cross-Correlation Method 
and Results}\label{S:shifts:results}

We applied the above method to identify variable velocity candidates
which we then scrutinized carefully to ensure that the shifts were
genuine. We rejected cases where substantial variations of the broad
line profiles mimic shifts (see example in
Figure~\ref{F:crosscorr}). We screened the initial candidates as
follows: First we checked that the result was the same regardless of
which of the two spectra was shifted and rebinned in the
cross-correlation procedure (if more than one followup observation was
carried out we also compared the results from the different
observations). Then we inspected the spectra to ensure that the shift
could be identified visually. In cases where the shift was subtle, we
experimented by varying the spectral windows over which the $\chi^2$
statistic was computed in order to verify that the result was robust.
Finally, for the candidates that passed the above tests, we carried
out simulations to verify the uncertainties: we made a copy of the
spectrum with the higher-$S/N$ added synthetic noise to match that of
the lower-$S/N$ spectrum and then applied the cross-correlation
algorithm. After 1000 realizations we examined the distribution of
shifts about zero to verify that the statistical uncertainty in the
shift was not underestimated. After arriving at the final statistical
uncertainty in the velocity change of each object, we added it in
quadrature with the uncertainty in aligning the SDSS and post-SDSS
spectra via the \fion{O}{3} lines (see the last paragraph of
\S\ref{S:reduction}) to obtain the total uncertainty.

\input  tab5.tex

In Table~\ref{T:crosscorr} we summarize the results of theta
cross-correlation analysis. In 14 cases the broad H$\beta$ peak
velocities change significantly (at 99\% confidence).  In 38 cases we
obtain only limits to possible shifts (at 99\% confidence).  In 10 of
these cases we encountered small variations in the broad H$\beta$
profiles, such as small changes in the widths or variations in one of
the two wings. Even though we do report shifts or limits in 
these cases, they are subject to an additional systematic error, which
we have included in the limits we report in
Table~\ref{T:crosscorr}. In the remaining 16 cases we are not able to
obtain meaningful measurements of velocity changes, or limits, either
because the broad H$\beta$ profiles varied substantially between
observations or because the $S/N$ of of one of the two spectra was
extremely low. We also note that \citet{shields09b} have determined 
an upper limit to the velocity change in J105041 of $31\pm60\kms$ 
(or $[-30, 90]$, in the notation of Table~\ref{T:crosscorr}).

\begin{figure}
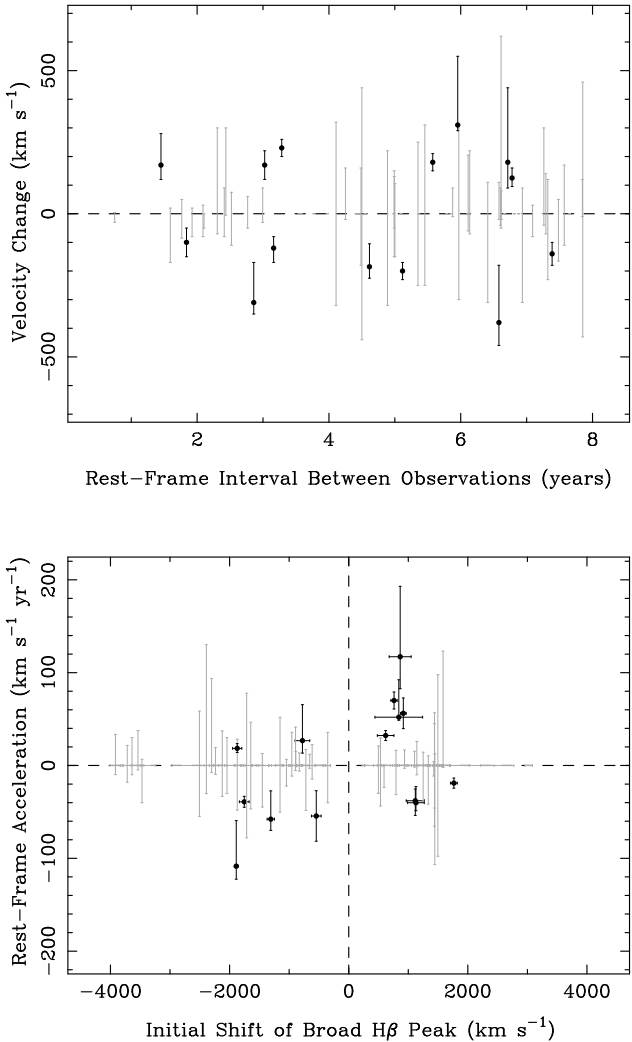

\centerline{\includegraphics[scale=0.4,angle=-90]{f10a.eps}}\bigskip \bigskip
\centerline{\includegraphics[scale=0.4,angle=-90]{f10b.eps}}
\caption{ {\it Top:} Velocity change of broad H$\beta$ peak ($\Delta
  v$) plotted against the \redtext{rest-frame} time interval between
  observations.  {\it Bottom}: Accelerations determined from changes
  in the broad H$\beta$ peak velocities ($\Delta v/\Delta t_{\rm
    rest}$) plotted against the initial peak velocity (peak offset, as
  measured from the original SDSS spectra). In both panels, the black
  points with error bars show the objects where a significant change
  in the broad H$\beta$ peak velocity was measured in the time
  interval covered by the observations. The light grey error bars show
  objects where only upper limits on the velocity change were obtained
  \citep[included are also the limits derived for J105041
    by][]{shields09b}. All error bars and limits correspond to the
  99\% ($2.6\sigma$) confidence level.
\label{F:accelerations}}
\end{figure}

In Figure~\ref{F:accelerations} we present our results graphically
\citep[including the limits on J105041, determined
  by][]{shields09b}. In the upper panel of this figure we plot the
measured velocity change against the \redtext{rest-frame} time
interval between observations. The statistically significant
measurements are denoted by black, solid circles with 99\%-confidence
error bars, while 99\%-confidence limits on the velocity changes are
shown as grey error bars.  An alternative representation of the
results is shown in the lower panel of the same figure where we plot
the observed acceleration against the initial shift of the broad
H$\beta$ peak (as measured from the SDSS spectra; see
Table~\ref{T:moments}).  We converted the velocity changes to
accelerations ($\Delta v/\Delta t_{\rm rest}$, in ${\rm
  km~s^{-1}~yr^{-1}}$) by dividing the velocity changes of
Table~\ref{T:crosscorr} by the \redtext{rest-frame} time intervals
between observations.  The velocity changes appear to be distributed
evenly with interval between observations. Moreover, there appears to
be no relation between the acceleration and the initial velocity
shift. We consider these findings further in our discussion of the
results in \S\ref{S:discussion}.

\subsection{Limitations and Caveats}\label{S:caveats}

As we noted above, there are a number of cases where the profile of
the broad H$\beta$ line has varied substantially between the SDSS and
post-SDSS observations (see Table~\ref{T:crosscorr}). Such an effect
represents an inevitable limitation to our ability to determine
changes in the velocity of the broad lines, which relies on the
long-term stability of the profiles. Variability of the line profiles
can occur on time scales comparable to the dynamical time of the BLR,
which should be shorter than the orbital period of the hypothesized
SBHB.  Examples of objects with substantially variable broad H$\beta$
profile include J112751, J113706, J140700, and J180545 (see the
comparison plots in Fig.~\ref{F:spectra}). A particularly interesting
example of substantial variability is J093653, where the broad
H$\beta$ line declined in flux between the two epochs of observation
with the result that the displaced peak of the broad line is now
blended with the narrow line, making the measurement of a velocity
change unreliable.

More subtle profile variations can also affect our ability to
determine velocity changes. For example, in J074157 (see
Fig.~\ref{F:crosscorr}) the broad H$\beta$ line became narrower
between the two observations and the line wings have moved closer to
the line core, which prevents us from obtaining limits on the
acceleration. In some cases, e.g., J091833, J111916, and J154340 (see
Fig.~\ref{F:spectra}), one of the two sides of the line has changed
but the other has not; we can determine limits on the acceleration in
such cases, but the uncertainties are larger than they would have been
if the profile were stable, as we indicate in Table~\ref{T:crosscorr}.

Yet another type of profile variability that can influence our results
is a correlated change on the two sides of the broad H$\beta$ profile
(a seesaw-like pattern) that can mimic a velocity change. Such
variability cycles have been observed in broad double-peaked emission
lines \citep[see, for example,][]{lewis10}. The beginnings of this
pattern are discernible in the spectra of J031715 and J160243 shown in
Figure~\ref{F:spectra}, for example. With this in mind, we note that
some of the objects for which we report statistically significant
velocity changes, specifically those where the shifted peak of the
broad H$\beta$ line is partially blended with the narrow H$\beta$
line, may be subject to this effect (e.g., J093844, J120924). Thus,
the velocity changes reported in Table~\ref{T:crosscorr} are subject
to verification by future followup observations. There are, of course,
additional reasons why further followup observations are needed, as we
discuss in more detail in the next section.

\section{Discussion}\label{S:discussion}

\subsection{Alternative Interpretations of Displaced Emission Line Peaks}\label{S:disc-alt}

Given that the starting point for our search for SBHBs and
rapidly-recoiling BHs was the selection of broad Balmer lines whose
peaks are displaced from their nominal wavelengths, additional
observations are needed to find the strong candidates from the initial
list.  The reason is that we cannot exclude alternative
interpretations for the displaced peaks of the broad lines such as an
unusual structure or a perturbation of the BLR.  For example, in the
class of models where the BLR is the surface of the accretion disk or
the base of the wind that is launched from it, perturbations in the
form of large-scale spiral arms can create line profiles with a
dominant peak or shoulder that is shifted from the nominal wavelength
of the line. Examples of model profiles with such properties can be
found in \citet[][e.g., their Figure~36]{lewis10}. In the same
context, one can also envision cases where the broad Balmer lines from
an unperturbed disk are double-peaked but a perturbation such as a
spiral arm causes one of the two peaks to be considerably stronger
than the other. Many examples of observed profiles that have such
shapes can be found in in the Appendix of \citet{gezari07}; some of
the more remarkable cases are reproduced in Figure~5 of
\citet{lauer09}. Therefore, it is essential to carry out monitoring
observations to look for further variations in the velocities of the
displaced peak of the broad line.

The detection of velocity changes in broad H$\beta$ lines between two
epochs is only the first step in making the case that Balmer lines
with displaced peaks are indeed signposts of SBHBs.  Many monitoring
observations are needed to verify that the velocity changes continue
monotonically and that they follow the pattern expected for orbital
motion. In particular the peaks of the broad line must drift from a
negative velocity to a positive velocity and {\it vice versa}.  Since
our understanding of the structure and dynamics of the gas in the BLR
is incomplete, we cannot select SBHBs without observing complete
orbital cycles because intrinsic variations of the broad line profiles
on time scales of several years may resemble the pattern of orbital
motion. A good case in point is provided by Mrk~668 (included in our
sample here as J140700). This object was one of the first two proposed
SBHBs \citep{gaskell83} and the variability of its broad Balmer lines
has been monitored over a long period of time \citep[$\sim 16$ years;
  see][]{marziani93,gezari07}. Although the velocity of the peak of
the Balmer lines does change sign, the radial velocity curve is not
compatible with orbital motion: the shape of the curve is not
sinusoidal, the wings of the line profile do not change accordingly
\citep[see Figures 30 and 31 of][]{gezari07}, and the center of mass
of a hypothesized binary appears to be at a different velocity than
the host galaxy \citep{marziani93}. This point is also bolstered by the
case of 3C~390.3, a double-peaked emitter whose H$\beta$ line changed
in a manner reminiscent of orbital motion by $\sim {1\over 4}$ of a
cycle \citep{gaskell96}.  The pattern changed drastically however
after that and became incompatible with orbital motion
\citep{eracleous97, shapovalova01}. More examples of broad Balmer
lines with displaced peaks whose velocities have varied in a manner
inconsistent with orbital motion can be found in \citet{gezari07} and
\citet{lewis10}.  The need for long-term monitoring makes the search
for SBHBs a challenging exercise in patience and persistence since the
expected orbital periods can be quite long (see
equation~[\ref{Q:separationnum}] and the associated discussion).

Notwithstanding the caveats listed above, we consider the consequences
of our findings for the two scenarios we have set out to test.  In
summary, the main observational results of this work that bear
directly on our tests are: (a) We have identified 88 quasars at $z\ls
0.7$ (among 15,900 quasars from SDSS DR7) whose broad H$\beta$ lines
are shifted from their nominal positions by a few$\times 1,000$\kms,
(b) we have measured statistically significant accelerations of the
offset peaks (in the range of \redtext{$-120$ to +120${\rm \; km\;
    s^{-1}\; yr^{-1}}$}) in 14/68 objects via followup observations,
(c) we have measured the properties of the broad H$\beta$ lines from
the SDSS spectra and found a correlation between the skewness of the
profiles and the velocity offset of the peak, which suggests that all
the objects of this sample belong to the same family, \bluetext{i.e.,
  have a single, common physical explanation} and (d) we have measured
the properties several narrow, forbidden, optical emission lines and
found them to be very similar to those of typical quasars.

\subsection{Implications for Rapidly Recoiling BHs}\label{S:disc-recoil}

\bluetext{The rapidly recoiling BH hypothesis, remains a viable
  possibility for objects in this sample.  This is because the primary
  observable, the velocity offset of the broad emission lines, can be
  interpreted both as a result of an SBHB and as a result of a large
  recoil speed. Observational tests that have been proposed to
  distinguish between the two interpretations are indirect.  The
  likelihood that a significant faction of our sample are rapidly
  recoiling black holes appears low a priori on theoretical
  grounds. In particular, \citet{dotti10} find that only 0.2\% of all
  recoil speeds should exceed 400\kms, while the distribution of line
  shifts that we find, if all of these are interpreted as recoil
  speeds, implies that $>0.5\%$ of recoil speeds exceed 1,000\kms.  In
  the cases where we were able to measure an acceleration, a rapidly
  recoiling BH interpretation is not ruled out because intrinsic
  variability of the line profiles can mimic an acceleration as we
  discuss in \S\ref{S:disc-alt}. However, if the accelerations we have
  measured \redtext{are confirmed to be real accelerations}, that
  would weaken the case for a rapidly recoiling BH and strengthen the
  SBHB interpretation for those particular objects.}

An observational selection criterion for recoiling BH candidates,
proposed by \citet{bonning07}, is based on the widths of the narrow,
forbidden lines.  If the displaced quasar is illuminating the gas in
its former host galaxy from the outside, the narrow-line region is not
stratified in ionization, thus the widths of all the narrow, forbidden
lines should be approximately the same. \bluetext{If, however, the gas
  emitting the narrow lines is associated with the recoiling BH
  \citep[it could originate, for example, from biconical outflows from
    the accretion disk, as suggested by][]{komossa08a}, the widths of
  the lines would not be a good indicator of such a system since they
  may display all the terends observed in Seyfert galaxies. In such a
  case the luminosities and relative intensities of the narrow
  emission lines may provide a better test for recoiling BHs that
  their widths.}

Nevertheless, in the 50 quasars for which we were able to measure the
FWHM of the \fion{Ne}{5}, \fion{Ne}{3}, and \fion{O}{3} lines, there
are 9 cases where the widths of these lines appear to be the same
within 10\%. In 8 of these cases, the \fion{O}{2} lines are
significantly narrower. The exception is J084313, where the width of
the \fion{O}{2} is comparable to that of the other forbidden lines. In
this object, the velocity offset of the broad H$\beta$ line from the
system of the narrow lines is only $-610\kms$ and the profile of the
line if fairly symmetric, as indicated by the skewness coefficients
\citep[this is another important consideration for selecting candidate
  recoiling BHs according to][]{shields09b}. In spite of the above,
the relative strengths of the narrow, forbidden lines of this object
are very typical of a Seyfert galaxy, which suggests that the
ionization parameter of the gas in the narrow-line region should be
similar to that of Seyferts. This is not in agreement with the
proposed picture of a displaced quasar that illuminates its host
galaxy from a significant distance. The quasar J092712, proposed as a
candidate rapidly recoiling BH by \citet{komossa08a}, has two systems
of narrow lines and in the one that would be attributed to the
(former) host galaxy in this interpretation the narrow lines are
unusually narrow and have similar widths. However, alternative
interpretations have also been proposed \citep{bogdanovic09, dotti09a,
  shields09a, heckman09}, \bluetext{which means that the rapidly
  recoiling BH scenario is not a unique interpretation for this very
  unusual object.}

\bluetext{We note in conclusion that the process of photoionization of
  the gas in a galaxy by an offset or neighboring AGN warrants further
  investigation in order to develop more quantitative and robust
  emission-line diagnostics. On the theoretical side, the
  photoionization calculations of \citet{gnedin97} can be extended to
  cover a wider range of conditions and to make predictions for the
  relative intensities, luminosities, and profiles of the narrow
  emission lines. On the observational side, progress can be made by
  seeking out and studying more analogs of the Was~49 system
  \citep{moran92}, a dual Seyfert galaxy in which the more luminous
  AGN contributes to the ionization of the interstellar medium of the
  host galaxy of its less luminous neighbor.}

\subsection{Implications for SBHBs}\label{S:disc-sbhb}

Considering our results in the context of the SBHB hypothesis, we can
compare them with the specific predictions made by \citet{volonteri09}
for the number of such systems that may be present in the SDSS DR7
quasar sample at $z<0.7$. If we suppose that all of the objects in our
sample are SBHBs, then their number is broadly consistent with one of
the two models considered by \citet{volonteri09}, their model
I. According to this model, which is based on the assumption that all
SBHBs are active (i.e., accreting) regardless of the merger history of
their host, \bluetext{there are at most 160 systems with $q >
  10^{-2}$} in the SDSS DR7 quasar sample. If we only take the 14
systems with detectable velocity changes to be SBHBs, their number is
broadly consistent with model II of \citet{volonteri09} according to
which SBHBs are active only in the aftermath of a major merger. Under
the assumptions of this model \bluetext{the predicted number of SBHBs
  with $q > 10^{-2}$ in the SDSS DR7 quasar sample is
  16}. \bluetext{The comparison of the above predictions with the
  observational results is complicated by a number of
  uncertainties. On the theoretical side: (a) the models refer to a
  volume-limited sample of quasars, while the SDSS quasar sample is
  not volume limited (given the flux limit of the quasar catalog,
  quasars with $i$ magnitudes between 19.1 and 21.2 are missed even
  though they would sattisfy the quasar luminosity criterion), (b) the
  predicted numbers quoted above rely on the assumption that both of
  the BHs in a binary may accrete (with an assumed distribution of
  Eddington ratios), without regard to their relative luminosity,
  while our selection method picks out systems in which the BLR around
  only one of the two BHs is visible, (c) inclusion in the theoretical
  sample is based on the luminosity of the primary, regardless of
  whether or not it is less luminous than the secondary. The above
  uncertainties can change the upper limits quoted above in either
  direction. On the observational side: (a) some of the candidates we
  report may turn out not to be SBHBs for the reasons discussed in
  \S\ref{S:disc-alt}, (b) if we assume that our selection method is
  appropriate for finding SBHBs, we are likely to have missed systems
  with small projected orbital velocities (i.e., systems viewed close
  to face on, or caught near conjunction, or with a combination of low
  masses and wide separations), and (c) the number of objects for
  which a measurable acceleration is likely to change change with
  future observations.}

Assuming that velocity changes we have measured correspond to orbital
accelerations in an SBHB we can explore the implications of the
results for this hypothesis. The correlation between the skewness of
the broad H$\beta$ profiles and the shift of their peaks (see
Figure~\ref{F:shift-skew}) suggests that all the objects in this
sample belong to the same family\footnote{This is an optimistic
  hypothesis. A more pessimistic interpretation of this correlation
  is, of course, that none of the objects in our sample are SBHBs.},
therefore we ask why an acceleration was measurable only in a small
fraction of the sample. As we noted in \S\ref{S:design} we expect that
the SBHBs we can detect with our strategy can have periods from a few
decades to a few centuries depending on the total mass of the binary
and the observed velocity shift, which constrains the orbital
separation given the masses of the two BHs. As pointed out by
\citet{loeb10}, the likelihood of finding a system with a given set of
intrinsic properties depends on its decay time via emission of
gravitational radiation. This can be estimated from the formula of
\cite{peters64}, which we can cast in terms of intrinsic system
properties and observables using equation (\ref{Q:separationnum}) as
follows:
\begin{eqnarray}
t_{\rm gr} 
& = & {5\over 256} {c^5 a^4\over (GM)^3} {(1+q)^2\over q}\;  \nonumber \\
& = & 4\times 10^{13}\, M_8\, {(1+q)^2\over q}\,
\left[\sin i\,|\sin\phi|\over(1+q)\,u_{2,3}\right]^8~{\rm yr} \label{Q:grnum} \\
& = & {8\times 10^{11} \, M_8 \over (1+q)^6\,u_{2,3}^8}
\left(0.1\over q\right)\, 
\left[
{\sin i \over \sin 45^\circ}\,{|\sin\phi|\over\sin 45^\circ}
\right]^8~{\rm yr}\, . \nonumber  \\
\nonumber 
\end{eqnarray}
Since the lifetime of an observed SBHB is such a sensitive function of
the observed velocity offset of the broad emission line, $u_{2,3}$, we
would expect that low-mass, high-speed binaries should be extremely
rare since their orbits should be rapidly decaying by gravitational
radiation. Specifically, setting $M_8=0.1$ and $u_{2,3}=2$ in
equation~(\ref{Q:grnum}) gives $t_{\rm gr}\approx 3\times 10^8$~yr,
while setting $M_8=1$ and $u_{2,3}=3$ gives $t_{\rm gr}\approx 1\times
10^8$~yr. The above considerations lead us to conclude that systems
with observed velocity shifts in excess of 1,000\kms\ may be (a)
highly inclined (hence with large separations, according to equation
[\ref{Q:separationnum}]), (b) caught near quadrature, (c) very
massive, or (d) very small mass-ratio systems ($q\ll 0.1$). Under
different conditions, such systems would have very short
lifetimes. The relatively uniform distribution of velocity changes and
accelerations depicted in Figure~\ref{F:accelerations} suggests that a
combination of the above four effects is at work. Highly inclined or
massive systems would have low accelerations, according to equation
(\ref{Q:accelerationnum}), which may explain why we were not able to
detect an acceleration in the majority of the objects we have followed
up.  Continued monitoring can discriminate among the above
possibilities by providing constraints on the binary phase through the
shape of the radial velocity curve.

Useful constraints on the properties of a hypothesized SBHB can be
obtained from a few observations spanning a baseline of about a
decade. Accurate measurements of the velocities of the peaks over such
a baseline can constrain the curvature of the radial velocity curve
and yield a lower limit on the period and total mass of the SBHB. This
technique, pioneered by \citet{halpern88} and developed further by
\citet{eracleous97}, was used to test and eventually reject the SBHB
hypothesis for three quasars with double-peaked Balmer lines. An
application of the same technique to objects from the present sample
can lead to similar constraints, which could render the SBHB
hypothesis untenable if these constraints are restrictive enough. To
this end, we are continuing to monitor all of our targets with the
immediate goal of collecting a few radial velocity measurements for
each object and using them to evaluate the plausibility of the SBHB
hypothesis by the above method.

Another avenue for testing the SBHB hypothesis for shifted broad
Balmer line peaks involves a comparison of the profiles of the
Ly$\alpha$ and Balmer lines. If the displaced peaks originate in a
perturbed disk around a single BH rather than in a BLR bound to one of
the two BHs in SBHB, we would expect the profiles of the broad optical
and UV emission lines to differ considerably. This expectation is
based on an analogy with nearby, well-studied double-peaked emitters
\citep[see, for example,][]{eracleous06, eracleous09}. The most
dramatic difference in such a case is between the profile of
Ly$\alpha$, which peaks at its nominal wavelength and those of the
Balmer lines whose peaks are displaced. We are currently pursuing this
test using UV spectra from the {\it Hubble Space Telescope}.

\acknowledgements

We are grateful to Steinn Sigurdsson for many stimulating discussions
and to Tamara Bogdanovi\'c for her critical reading of the manuscript.
\bluetext{We also acknowledge helpful discussions with Marta Volonteri
  and Cole Miller. M.E. gratefully acknowledges the hospitality of the
  Aspen Center for Physics where the final stages of this work were
  carried out.}

The Hobby-Eberly Telescope (HET) is a joint project of the University
of Texas at Austin, the Pennsylvania State University, Stanford
University, Ludwig-`Maximillians-Universit\"at M\"unchen, and
Georg-August-Universit\"at G\"ottingen. The HET is named in honor of
its principal benefactors, William P. Hobby and Robert E. Eberly.

The Marcario Low-Resolution Spectrograph is named for Mike Marcario of
High Lonesome Optics, who fabricated several optics for the instrument
but died before its completion; it is a joint project of the
Hobby-Eberly Telescope partnership and the Instituto de
Astronom\'{\i}a de la Universidad Nacional Aut\'onoma de M\'exico.

This research has made use of the NASA/IPAC Extragalactic Database
(NED) which is operated by the Jet Propulsion Laboratory, California
Institute of Technology, under contract with the National Aeronautics
and Space Administration.


\input references.tex
\end{document}

%% file: tab1.tex
\begin{deluxetable*}{lccccccccc}
\tablewidth{0in}
\tabletypesize{\scriptsize}
\tablecolumns{10}
\tablecaption{Log of Spectroscopic Observations\label{T:obslog}}
\tablehead{
\colhead{} &
\colhead{} &
\colhead{} &
\colhead{} &
\colhead{} &
\colhead{} &
\colhead{} &
\colhead{Exposure} &
\colhead{} &
\colhead{Rest-Frame} \\
\colhead{Object Name} &
\colhead{} &
\colhead{$m_{\rm V}$\tablenotemark{b}} &
\colhead{$A_{\rm V}$\tablenotemark{c}} &
\colhead{$M_{\rm V}$\tablenotemark{d}} &
\colhead{Observation} &
\colhead{Instr.} &
\colhead{Time} &
\colhead{} &
\colhead{Wavelength} \\
\colhead{SDSS J} &
\colhead{$z$\tablenotemark{a}} &
\colhead{(mag)} &
\colhead{(mag)} &
\colhead{(mag)} &
\colhead{Date (UT)} &
\colhead{Code\tablenotemark{e}} &
\colhead{(s)} &
\colhead{$S/N$\tablenotemark{f}} &
\colhead{Range (\AA)} \\
\colhead{(1)} &
\colhead{(2)} &
\colhead{(3)} &
\colhead{(4)} &
\colhead{(5)} &
\colhead{(6)} &
\colhead{(7)} &
\colhead{(8)} &
\colhead{(9)} &
\colhead{(10)} 
}
\startdata
\\
$001224.02-102226.2$ & 0.2287 & 17.07 &  0.127 & --23.27 & 2001/08/20 & S  &  3756 &   29 & 3093--7505 \\ 
                     &        &       &        &         & 2009/12/16 & M  &  3600 &   28 & 3181--6165 \\ 
\\
$002444.10+003221.3$ & 0.4024 & 16.85 &  0.083 & --24.86 & 2000/12/22 & S  &  9900 &   88 & 2709--6594 \\ 
                     &        &       &        &         & 2009/12/18 & M  &  2400 &   30 & 2811--5426 \\ 
\\
$015530.01-085704.0$ & 0.1648 & 16.84 &  0.080 & --22.65 & 2001/09/16 & S  &  2702 &   32 & 3263--7917 \\ 
                     &        &       &        &         & 2009/12/17 & M  &  3600 &   55 & 3375--6524 \\ 
\\
$020011.53-093126.2$ & 0.3602 & 17.96 &  0.081 & --23.46 & 2001/08/28 & S  &  8104 &   34 & 2795--6780 \\ 
                     &        &       &        &         & 2009/12/17 & M  &  3600 &   28 & 2890--5586 \\ 
\\
$021259.59-003029.5$ & 0.3945 & 17.87 &  0.117 & --23.83 & 2000/09/29 & S  &  2700 &   29 & 2714--6613 \\ 
                     &        &       &        &         & 2009/12/16 & M  &  3600 &   13 & 2803--5433 \\ 
\\
$022014.57-072859.1$ & 0.2137 & 18.65 &  0.078 & --21.47 & 2001/09/10 & S  &  4501 &   11 & 3133--7598 \\ 
                     &        &       &        &         & 2009/12/16 & M  &  3600 &   16 & 3220--6241 \\ 
\multicolumn{10}{c}{\hfill} \\
\multicolumn{10}{c}{The remaining objects have been omitted} \\
\multicolumn{10}{c}{The full table can be found in the complete preprint} \\
\enddata
\tablenotetext{a}{The redshift of the object as measured in this paper
  from the peak wavelenth of [\ion{O}{3}]~$\lambda$5007 line.}
\tablenotetext{b}{The apparent V magnitude, determined from the SDSS
  PSF magnitudes, as described in \S\ref{S:sample-properties} of the
  text.}
\tablenotetext{c}{The Galactic visual extinction, taken from
  \citet*{schlegel98}}
\tablenotetext{d}{The absolute V magnitude, computed as described in
  \S\ref{S:sample-properties} of the text.}
\tablenotetext{e}{The telescope and instrument configuration, as
  described in Table~\ref{T:telcodes}.}
\tablenotetext{f}{The signal-to-noise ratio ($S/N$) in the continuum
  near the line of interest.  When the spectrum includes the H$\beta$
  line we give the $S/N$ in the continuum near this line, at 4600~\AA.
  If the spectrum includes only the \ion{Mg}{2}~$\lambda$2800 line, we
  give the $S/N$ in the continuum near this line, at 2900~\AA.}
\end{deluxetable*}

%% file: tab2.tex
\begin{deluxetable*}{cllc}
\tablewidth{0in}
\tabletypesize{\scriptsize}
\tablecolumns{5}
\tablecaption{List of Telescopes and Instruments\label{T:telcodes}}
\tablehead{
\colhead{Instrument} &
\colhead{} &
\colhead{} &
\colhead{Resolution\tablenotemark{a}} \\
\colhead{Code} &
\colhead{Observatory, Telescope, and Spectrograph} &
\colhead{Spectral Elements} &
\colhead{(\AA)} 
}
\startdata
S  & SDSS 2.5m, SDSS spectrograph                          & 640, 440 mm$^{-1}$ grisms, $d=3$\arcsec\ fibers & 2.7 \\
\\
M  & MDM, Hiltner 2.4m, Boller \& Chivens CCD spectrograph & G350L grating (150 mm$^{-1}$), 1\farcs0 slit           & 7.6 \\
\\
K  & KPNO, Mayall 4m, Ritchie-Cretien spectrograph         & KPC-007 grating (600 mm$^{-1}$), 1\farcs5 slit &  3.1 \\
\\
Pb & Palomar, Hale 5m, Double Spectrograph                 & blue arm: 600 mm$^{-1}$ grating, 1\farcs5 slit &  4.0 \\
Pr &                                                       & red arm: 600 mm$^{-1}$ grating, 1\farcs5 slit  &  3.1 \\
\\
H1 & Hobby-Eberly Telescope, Low-Resolution Spectrograph   & G1 grism (300 mm$^{-1}$), 1\farcs5 slit        & 14.5  \\
H2 &                                                       & G2 grism (600 mm$^{-1}$), 1\farcs5 slit        &  5.6 \\
\enddata
\end{deluxetable*}

%% file: tab3.tex
\def\tt#1{$\times 10^{#1}$}
\def\un{\tablenotemark{e}}

\begin{deluxetable*}{crrrrcrrrrcccccc}
\tablewidth{7.in}
\tabletypesize{\scriptsize}
\tablecolumns{16}
\tablecaption{Properties of Narrow Emission Lines\label{T:narrow}}
\tablehead{
\colhead{} &
\multicolumn{4}{c}{Obs. Flux Relative to \fion{O}{3}\tablenotemark{a}} &
\colhead{} &
\multicolumn{4}{c}{Corr. Flux Relative to \fion{O}{3}\tablenotemark{b}} &
\colhead{} &
\colhead{} &
\multicolumn{4}{c}{FWHM (km s$^{-1}$)\tablenotemark{d}} \\
\colhead{} &
\multicolumn{4}{c}{\hrulefill} &
\colhead{Obs.} &
\multicolumn{4}{c}{\hrulefill} &
\colhead{Corr.} &
\colhead{} &
\multicolumn{4}{c}{\hrulefill} \\
\colhead{Object} &
\colhead{\fion{Ne}{5}} &
\colhead{\fion{O}{2}}  &
\colhead{\fion{Ne}{3}} &
\colhead{H$\beta$}     &
\colhead{$F_{\rm [O\,III]}$\tablenotemark{a}}  &
\colhead{\fion{Ne}{5}} &
\colhead{\fion{O}{2}}  &
\colhead{\fion{Ne}{3}} &
\colhead{H$\beta$}     &
\colhead{$F_{\rm [O\,III]}$\tablenotemark{b}}  &
\colhead{$L_{\rm [O\,III]}$\tablenotemark{c}}  &
\colhead{\fion{Ne}{5}} &
\colhead{\fion{Ne}{3}} &
\colhead{H$\beta$}     &
\colhead{\fion{O}{3}}  \\
\colhead{(1)} &
\colhead{(2)} &
\colhead{(3)} &
\colhead{(4)} &
\colhead{(5)} &
\colhead{(6)} &
\colhead{(7)} &
\colhead{(8)} &
\colhead{(9)} &
\colhead{(10)} &
\colhead{(11)} &
\colhead{(12)} &
\colhead{(13)} &
\colhead{(14)} &
\colhead{(15)} &
\colhead{(16)} 
}
\startdata
J001224  &$<$0.055   &   0.462   &   0.346   &   0.128   & 3.59 &$<$0.058   &   0.483   &   0.359   &   0.129   & 3.95 & 5.73 & ... & 910 & 510 & 540 \\
J002444  &   0.200   &   0.130   &   0.242   &   0.193   & 7.77 &   0.207   &   0.133   &   0.247   &   0.194   & 8.19 & 44.0 & 890 & 910 & 420 & 420 \\
J015530  &   0.157   &   0.166   &   0.184   &   0.237   & 16.9 &   0.162   &   0.171   &   0.189   &   0.238   & 18.0 & 12.5 & 670 & 550 & 470 & 290 \\
J020011  &$<$0.105   &   0.140   &   0.220   &   0.215   & 3.58 &$<$0.109   &   0.143   &   0.225   &   0.215   & 3.77 & 15.5 & ... &1150 & 720 & 800 \\
J021259  &   0.054   &   0.132   &   0.086   &   0.075   & 31.5 &   0.056   &   0.137   &   0.089   &   0.075   & 34.0 & 174. & 430 & 490 & 450 & 440 \\
J022014  &$<$0.112   &   0.564   &   0.151   &   0.219   & 3.90 &$<$0.116   &   0.579   &   0.155   &   0.220   & 4.14 & 5.14 & ... & 560 & 400 & 370 \\
\multicolumn{16}{c}{\hfill} \\
\multicolumn{16}{c}{The remaining objects have been omitted} \\
\enddata
\tablenotetext{a}{Integrated line fluxes are expressed relative to the 
integrated flux of the [\ion{O}{3}]~$\lambda$5007 line in units of
$10^{-15}{\rm erg\;cm^{-2}\;s^{-1}}$, as measured from the observed
spectrum, without any corrections. The uncertainty is 5\% or less,
unless otherwoise noted.}
\tablenotetext{b}{Integrated line fluxes are expressed relative to the 
integrated flux of the [\ion{O}{3}]~$\lambda$5007 line 
 in units of $10^{-15}{\rm erg\;cm^{-2}\;s^{-1}}$, after
correcting for Galactic extinction, as described
in \S\ref{S:sdss-lines} of the text. The uncertainty is 5\% or less,
unless otherwoise noted.}
\tablenotetext{c}{The luminosity of the [\ion{O}{3}]~$\lambda$5007 line
 in units of $10^{37}$\ergs, 
 after correcting for Galactic extinction (see \S\ref{S:sdss-lines} of
 the text).}
\tablenotetext{d}{The full width at half maximum of the emission lines,
corrected for the finite resolution of the spectrograph as described
in \S\ref{S:sdss-lines} of the text. The value is omitted when the line
is not detected.}
\tablenotetext{e}{The uncertainty in these measurements is 10--15\%}
\tablenotetext{f}{The spectrum of J092712 includes two sets of narrow lines therefore we report te 
properties of each set in separate rows. The first row gives the properties of the set that is 
blueshifted and appears to be at the same redshift as the broad lines. The second row gives the 
properties of the set that is reshifted and has no broad lines associated with it.}
\end{deluxetable*}

%% file: tab4.tex
\def\hb{\hbox to 0.5em{\hss}}
\def\bk{\!\!\!\!}

\begin{deluxetable*}{crrrlrcrr}
\tablewidth{0in}
\tabletypesize{\scriptsize}
\tablecolumns{9}
\tablecaption{Properties of Broad H$\beta$ Profiles\label{T:moments}}
\tablehead{
\colhead{} &
\colhead{} &
\colhead{} &
\multicolumn{2}{c}{Peak} &
\colhead{Centroid} &
\colhead{Velocity} &
\colhead{} &
\colhead{Pearson} \\
\colhead{} &
\colhead{FWHM} &
\colhead{FWQM} &
\multicolumn{2}{c}{Velocity Shift} &
\colhead{Velocity Shift} &
\colhead{Dispersion} &
\colhead{Skewness} &
\colhead{Skewness} \\
\colhead{Object} &
\colhead{(km s$^{-1}$)} &
\colhead{(km s$^{-1}$)} &
\multicolumn{2}{c}{(km s$^{-1}$)} &
\colhead{(km s$^{-1}$)} &
\colhead{(km s$^{-1}$)} &
\colhead{Coefficient} &
\colhead{Coefficient} \\
\colhead{(1)} &
\colhead{(2)} &
\colhead{(3)} &
\multicolumn{2}{c}{(4)} &
\colhead{(5)} &
\colhead{(6)} &
\colhead{(7)} &
\colhead{(8)} 
}
\startdata
 J001224 &  3590 \hb &   8630 \hb & --1870 & $\bk\pm$  80 &    280 \hb &  3530 &   0.253 \hb & --0.197 \hb \\
 J002444 &  9320 \hb &  13450 \hb &  --720 & $\bk\pm$ 150 &   --80 \hb &  3720 &   0.113 \hb & --0.056 \hb \\
 J015530 &  7320 \hb &   9180 \hb &   1430 & $\bk\pm$ 130 &    610 \hb &  3100 & --0.039 \hb &   0.045 \hb \\
 J020011 &  7660 \hb &  12490 \hb &   1650 & $\bk\pm$  60 &    460 \hb &  3570 &   0.333 \hb &   0.025 \hb \\
 J021259 &  7990 \hb &  11570 \hb & --2300 & $\bk\pm$  40 &   --10 \hb &  3830 &   0.323 \hb & --0.106 \hb \\
 J022014 &  8210 \hb &  10560 \hb &  --470 & $\bk\pm$  90 &    950 \hb &  3240 &   0.495 \hb & --0.102 \hb \\
\multicolumn{9}{c}{\hfill} \\
\multicolumn{9}{c}{The remaining objects have been omitted} \\
\enddata
\tablenotetext{a}{The redshift of the object as measured in this paper
  from the peak wavelenth of [\ion{O}{3}]~$\lambda$5007 line.}
\tablenotetext{b}{The apparent V magnitude, determined from the SDSS
  PSF magnitudes, as described in \S\ref{S:sample-properties} of the
  text.}
\tablenotetext{c}{The Galactic visual extinction, taken from
  \citet*{schlegel98}}
\tablenotetext{d}{The absolute V magnitude, computed as described in
  \S\ref{S:sample-properties} of the text.}
\tablenotetext{e}{The telescope and instrument configuration, as
  described in Table~\ref{T:telcodes}.}
\tablenotetext{f}{The signal-to-noise ratio ($S/N$) in the continuum
  near the line of interest.  When the spectrum includes the H$\beta$
  line we give the $S/N$ in the continuum near this line, at 4600~\AA.
  If the spectrum includes only the \ion{Mg}{2}~$\lambda$2800 line, we
  give the $S/N$ in the continuum near this line, at 2900~\AA.}
\end{deluxetable*}

%% file: tab5.tex
\def\fnb{\tablenotemark{\;{\rm b}}}
\def\fnc{\tablenotemark{\;{\rm c}}}
\def\fnd{\tablenotemark{\;{\rm d}}}
\def\marg{}
\begin{deluxetable*}{ccccccccc}
\tablewidth{0in}
\tabletypesize{\small}
\tablecolumns{7}
\tablecaption{Reasults of Cross-Correlation Analysis\label{T:crosscorr}}
\tablehead{
\colhead{} & \colhead{Broad H$\beta$ Shift\tablenotemark{a}} & \colhead{\marg} & \colhead{} & \colhead{Broad H$\beta$ Shift\tablenotemark{a}} & \colhead{\marg} & 
\colhead{} & \colhead{Broad H$\beta$ Shift\tablenotemark{a}} & \colhead{} \\
\colhead{Object} & \colhead{(km s$^{-1}$)} & \colhead{} & \colhead{Object} & \colhead{(km s$^{-1}$)} & \colhead{} & 
\colhead{Object} & \colhead{(km s$^{-1}$)} & \colhead{} 
}
\startdata
J001224 & $+125^{+ 35}_{- 30}$   & \phantom{026} &  J102106 & $[-250,+310]$          & \phantom{057} &  J141300 & \dots$^{\rm d}$            & \phantom{056}  \\ 
J002444 & $[-310,+110]$          & \phantom{070} &  J104132 & \dots$^{\rm d}$            & \phantom{042} &  J143123 & \dots$^{\rm b}$            & \phantom{006}  \\ 
\multicolumn{8}{c}{\hfill} \\
\multicolumn{8}{c}{The remaining objects have been omitted} \\
\enddata
\tablenotetext{a}{Numbers with error bars denote a statistically significant 
  shift. Pairs of numbers in square brackets denote upper and lower limits on 
  the shift. A positive shift means that the post-SDSS spectrum is shifted towards
  longer wavelengths.}
\tablenotetext{b}{The broad H$\beta$ profile varied substantially
  between observations, thus a meaningful shift could not be
  determined. These variaitons may include large changes in the
  profile shape or changes in the width that cause the two wings of
  the line to move in opposite directions.}
\tablenotetext{c}{Variability of the broad H$\beta$ line profile affects 
our ability to determine a shift and increases the uncertainties.}
\tablenotetext{d}{The low $S/N$ in the followup spectrum does not 
allow us to get meaningful constraints on the shift.}
\end{deluxetable*}